\documentclass [12pt]{article}
\usepackage{lscape}                                           %
\usepackage[centertags]{amsmath}
\usepackage{amsfonts} \usepackage{amssymb}
\usepackage{color}
\usepackage{graphicx}
\usepackage{lscape}
\newcommand{\bZ}{{\bar Z}}

\newcommand{\bn}{{\bar n}}
\newcommand{\bt}{{\bar t}}
\newcommand{\bu}{{\bar u}}

\newcommand{\bz}{{\bar z}}

\newcommand{\cB}{{\cal B}}

\newcommand{\cE}{{\cal E}}

\newcommand{\cN}{{\cal N}}

\newcommand{\cR}{{\cal R}}
\newcommand{\cU}{{\cal U}}

\newcommand{\cZ}{{\cal Z}}
\newcommand{\cbZ}{{\tilde{\cal Z}}}

\newcommand{\uno}{\mathbb{I}}

\newcommand{\N}{\mathbb{N}}
\newcommand{\Z}{\mathbb{Z}}

\newcommand{\R}{\mathbb{R}}
\newcommand{\C}{\mathbb{C}}

\newcommand{\Mon}[1]{ M_{(#1)} }
\newcommand{\tMon}[1]{\hat{M}_{(\bt, #1)}}

\newcommand{\NB}{{\hat N_B}}

\newcommand{\omegat}{ {\omega_{(\bt)}}{} }

\newcommand{\bomegat}{ {{\bar \omega}_{(\bt)}}{} }
\newcommand{\homegat}{ {{\hat \omega}_{(\bt)}}{} }
\newcommand{\bomega}{ {{\bar \omega} }{} }

\newcommand{\Ltr}[3]{#1_{(#2)\, #3} }
\newcommand{\Vd}[1]{\Ltr{V^\dagger}{#1}{ } }
\newcommand{\Vdr}[2]{\Ltr{V^\dagger}{#1}{#2} }
\newcommand{\tVd}[1]{ \Ltr{ {\tilde V}^\dagger}{#1}{} }
\newcommand{\tVdr}[2]{\Ltr{{\tilde V}^\dagger} {#1}{#2} }

\newcommand{\V}[1]{\Ltr{V}{#1}{}}
\newcommand{\tV}[1]{ \Ltr{{\tilde V}}{#1} }
\newcommand{\Vr}[2]{\Ltr{ V} {#1}{#2} }
\newcommand{\tVr}[2]{\Ltr{{\tilde V}} {#1}{#2} }

\newcommand{\du}{\partial_u}
\newcommand{\dub}{ {\bar\partial} _ {\bar u} }
\newcommand{\dz}{\partial}

\newcommand{\vect}[2]{\left( \begin{array}{c} #1 \\ #2 \end{array} \right) }

\newcommand{\mat}[4]{\left( \begin{array}{c c} #1 & #2\\ #3 & #4 \end{array} \right) }

\newcommand{\oh}{\frac{1}{2}}

\newcommand{\hyppar}[1]{{\bf [ #1 ] }}
\newcommand{\hypbas}[1]{{\bf \{ #1 \} }}

\newcommand{\dzero}[1]{d_{\hypbas{0} #1}}

\newcommand{\FF}[4]{
F\left( \begin{array}{c c} #1 & #2\\ \multicolumn{2}{c}{#3}  \end{array} ; #4 \right) 
}

\newcommand{\PP}[9]{
P\left\{ \begin{array}{c c c c} 
#1 & #4 & #7 & \\   
#2 & #5 & #8 & z\\   
#3 & #6 & #9 &\\   
\end{array} \right\} 
}

\newcommand{\PPom}[9]{
P\left\{ \begin{array}{c c c c} 
#1 & #4 & #7 & \\   
#2 & #5 & #8 & \omegat\\   
#3 & #6 & #9 &\\   
\end{array} \right\} 
}

\newcommand{\COMMENTO}[1]{}

\newcommand{\COMMENTOO}[1]{}

\begin{document}
\title{
Towards a fully stringy computation of Yukawa couplings
on non factorized tori and non abelian twist correlators (I):
the classical solution and action
}

\author{
{Igor Pesando$^1$}
\\
~\\
~\\
$^1$Dipartimento di Fisica, Universit\`a di Torino\\
and I.N.F.N. - sezione di Torino \\
Via P. Giuria 1, I-10125 Torino, Italy\\
\vspace{0.3cm}
\\{ipesando@to.infn.it}
}

\maketitle
\thispagestyle{empty}

\abstract{
We  consider the simplest possible setting of non abelian twist fields
which corresponds to $SU(2)$ monodromies.
We first review the theory of hypergeometric function and of the
solutions of the most general Fuchsian second order equation with
three singularities.
Then we solve the problem of writing the  general solution with
prescribed $U(2)$ monodromies.
We use this result to
compute the classical string solution corresponding to three $D2$ branes
in $\R^4$.
Despite the fact the configuration is supersymmetric the classical
string solution is not holomorphic.
Using the equation of motion and not the KLT approach 
we give a very simple expression for the classical action of the string.
We find that the classical action is not proportional to the area of
the triangle determined by the branes intersection points  
since the solution is not holomorphic 
.
Phenomenologically this means that the Yukawa couplings for these
supersymmetric configurations on non factorized tori 
are suppressed with respect to the factorized case.
}
\\
keywords:{ D-branes, Conformal Field Theory, Yukawa couplings}

\newpage

\section{Introduction and conclusions}

Since the beginning, D-branes have been very important in the formal
development of string theory as well as in attempts to apply string
theory to particle phenomenology and cosmology.
However, the requirement of chirality in any physically realistic
model  leads to a somewhat restricted number of possible D-brane
set-ups. 
An important class of models are intersecting brane models where
chiral fermions can arise at the intersection of two branes at angles.
Most of these computable models are based on $D6$ branes at angles in $T^6$ or
its orbifolds.

To ascertain the phenomenological viability of a model the computation
of Yukawa couplings and flavor changing neutral currents plays an
important role.  
This kind of computations involves the computations of (excited) twist
fields correlators.
Besides the previous computations many other computations often
involve correlators of twist fields and excited twist fields.
It is therefore important and interesting in its own to be able to
compute these correlators.
The literature concerning orbifolds (see for example
\cite{Dixon:1986qv}, \cite{orbifold_old}, \cite{orbifold_new})
intersecting D-branes on factorized tori (see for example
\cite{angles}), magnetic branes with commuting magnetic fluxes (see
for example \cite{magnetic}) or involving ``abelian'' twist fields in
various applications (see for example \cite{more_twists}) is very
vast.
These results are mainly based on the so called stress-tensor method
\cite{Dixon:1986qv} and concerns mainly non excited twists even if
results for excited twists \cite{excited_twists} were obtained.
Some of the previous results were also obtained in the infinite charge
formalism and boundary state formalism \cite{boundary}. 
Within the Reggeon framework (see for example \cite{SDS},
\cite{reggeon}) the generating functions for the three points
correlators were also obtained in a somewhat complex way.
Finally in \cite{Pesando:2014owa} and \cite{Pesando:2011ce} based on
previous results \cite{miei_prima} and a mixture of the path integral
approach with the reggeon approach the generating function of all the
correlators with an arbitrary number of (excited) twist fields and
usual vertices was given in the case of abelian twist fields.
These computations boil down to the knowledge of the Green function in
presence of twist fields and of the correlators of the plain twist
fields. In this way the computations were made systematic differently
from many previous papers where correlators with excited twisted
fields have been computed on a case by case basis without a clear
global picture.
The same results were then recovered using the canonical quantization
approach in \cite{Pesando:2014sca}.

Until now only the case of factorized tori has been considered at the
stringy level.
It is clear that the non factorized case is more generic and
technically by far more complex.
It concerns the so-called non abelian twists for which only a handful
papers can be found in the literature of the last 30 years \cite{non_abelian}.
It is therefore interesting to try to understand how special the results
from the factorized case are and to try to clarify the technical
issues involved. 

In this very technical paper we start the investigation of these
configurations.
We start considering the case of three $D6$ branes embedded in
$\R^{10}$.
The relevant configuration can be effectively described by three
euclidean $E2$ branes in $\C^2=\R^4$.
We can think of embedding the first $E2$ brane as $\Im Z^1=\Im Z^2=0$.
Then the second and third $E2$ branes are generically characterized by
a $SO(4)$ matrix (or more precisely by an equivalence class, i.e a
point in the Grassmannian $SO(4)/ SO(2) \times SO(2)$) which describes
how they are embedded with respect to the first one.
However we limit our analysis to the simplest case where these matrices are
characterized by an equivalence class of $SU(2)$ .
If these two matrices commute then we are in the abelian case if not
we deal with the by far more difficult non abelian case.
Even if we do not consider the most general case it is however
interesting enough to start grasping the issues involved.
Moreover this configuration is supersymmetric since there are spinors
invariant under the other $SU(2)$ of the ``internal'' rotation
$SO(4)\equiv SU(2) \times SU(2)$.

Due  to the technicality of the computations involved we have
preferred to write down the details therefore the paper has grown in
dimension making necessary to split it into different parts.
In this part we recapitulate the mathematical tools necessary and we
find the classical solution of the bosonic string.
In a companion paper we deal with the Green function which is
necessary to compute the correlators involving excited twist fields.
We are nevertheless still not very close to determine from first
principles the normalization of the three twist field correlator as
it happens also for all the other papers on the subject
\cite{non_abelian}. 
The reason being the impossibility of writing explicitly the classical
solution of the string with four branes with a non abelian
configuration.

We can nevertheless draw some interesting conclusions.
In particular using the path integral approach
the $\NB$ point twist field correlator can be written roughly as
\begin{align*}
\langle
\sigma_{M_1}(x_1) \dots \sigma_{M_\NB}(x_\NB)
\rangle
=
\cN( \{x_t,\, M_t \}_{1\le t \le \NB} )
 e^{-S_{E,cl}( \{ x_t,\, M_t \}_{1\le t \le \NB} ) }
,
\end{align*}
where $M_t$ with $1\le t \le \NB$ are the monodromies.
Therefore the knowledge of the classical solution gives the main
contribution $e^{-S_{E,cl}( \{ x_t,\, M_t \}_{1\le t \le \NB} ) }$
even if the quantum contribution $\cN( \{x_t,\, M_t \}_{1\le t \le
\NB} ) $ is necessary for the complete result.
It then follows that given three $D6_t$ ($1 \le t \le 3)$ branes the
leading order of the Yukawa coupling in a truely stringy computation
is given by
\begin{align*}
Y_{1 2 3} \propto e^{-S_{E,cl}( \{ x_t,\, M_t \}_{1\le t \le 3} ) }
.
\end{align*}
Naively one could think that $S_{E,cl}$ is simply the area of the
triangle determined by the three interaction points but it is not so.
These interaction points always define a 2 dimensional real plane in
$\R^4$ but differently from the cases discussed before in the literature
the embedding of the string worldsheet which follows from the equation
of motion is not  a flat triangle, i.e. a triangle which lies in the plane
determined by the three interactions points.
In fact figure \ref{fig:Endpoint} shows the actual line traced 
by the endpoint of the classical string while the naive path should be
a segment.
This implies that Yukawa couplings in non factorized models are
supressed with respect to the factorized ones.
The reason is that the classical string solution is not holomorphic
(this must not be confused with the fact that the branes emdeddings
are holomorphic in the proper set of coordinates).

The paper is organized as follows.
In section \ref{sect:monodromies} we recapitulate the classical
mathematics needed for the computation.
In particular we consider the monodromies associated with the general
solution of the second order Fuchsian equation with three singular
points located at $0$, $1$ and $\infty$.
Given the relation between the parameters of the equation and the
monodromies we solve the inverse problem, i.e. given the monodromies
in $U(2)$ find the properly normalized combination of solutions which
has the desired monodromy set.
This solution is obviously expressed using the hypergeometric function
as in eq.s (\ref{Basis0}, \ref{u2-to-abcdf}, \ref{d0-ratio}).
Another not so commonly appreciated result of the discussion is that
monodromies depend on whether the base point is the upper or lower
half plane.

In section \ref{sect:string-action-branes-conf} we consider the string
action and the boundary conditions we have to impose.
The boundary conditions are better expressed as a local problem for
the monodromies  on the double string coordinates and a global
problem.

In section \ref{sect:the_classical_solution} 
we then proceed to find the actual classical solution. 
The upshot of this is more general than the three D-branes case.
It turns out that for the $U(2)$ case 
the bulk of most of the information is
contained in the local behavior of the solution, the indeces of the
Fuchsian equation, which are determined by the modulus of the vector
$\vec n$ which parametrizes the $SU(2)$ monodromy as $M=\exp{(i 2 \pi
  \vec n \cdot \vec \sigma)}$.
The normalization of the solution  depends also on the other parameters.
Nevertheless the indeces are not sufficient to completely fix the
solution but in the simplest case we consider since there are the
accessory parameters.

Finally, in section \ref{sect:the_classical_action} we compute the
classical action corresponding to the the solution found.
We do this in a more general way which allows us express the action as
a linear function of some coefficients opposed to the usual way of
getting an expression quadratic.
Moreover we clearly show that in the holomorphic case the action has a
geometrical meaning.

\section{Monodromies of the hypergeometric function}
\label{sect:monodromies}
To solve the problem of finding the string classical solution, 
we are interested in finding complex functions
with a given set of singular points and monodromies.
A good starting point to construct these functions is to consider the
Fuchsian linear differential equations (which are reviewed in appendix
\ref {app:fuchsian}) since the solutions come naturally in vectors with
given monodromies.

Since we are interested in the case with three singular points and
$U(2)$ monodromies we would now like to summarize some basics facts on the
hypergeometric function which we need in the following.

Our main interest is the derivation of monodromies.
In particular we discuss one point which seems to be overlooked or implicit in the
literature, i.e. the monodromies do depend on the point we start the
loop\footnote{
Historically we notice that even the first paper on the monodromies for the
hypergeometric function by Riemann in 1857 \cite{Riemann1857} seems not
to consider the two cases.
}.
It is in fact well known that the homotopy group is defined starting
from a base point and that all of these groups are isomorphic.
This does not however mean that their representations in the vector
space of the solutions of the hypergeometric  equation are equal. 
And actually they are not.
In appendix \ref{app:disc2mon} we show this point in a local setup and
at the end of section \ref{sec:StringBoundaryConditions} we
explicitly show how this fact is needed to demonstrate that the
string action is well defined when it is written using the string
coordinates obtained by the doubling trick. 

\subsection{Paths}

\begin{figure}[hbt]
\begin{center}
\def\svgwidth{250px}
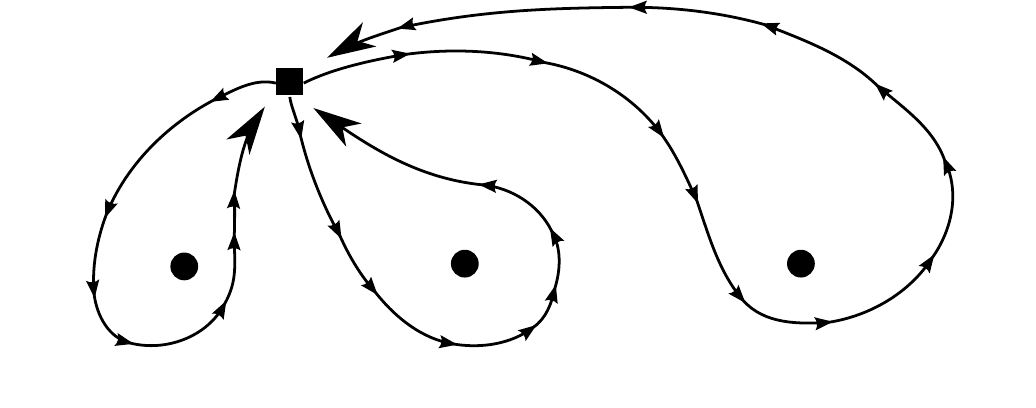
\end{center}
\vskip -0.5cm
\caption{The three different paths around the marked points
  $0,1,\infty$ starting in the upper half plane.
}
\label{fig:paths_in_H+}
\end{figure}

\begin{figure}[hbt]
\begin{center}
\def\svgwidth{250px}
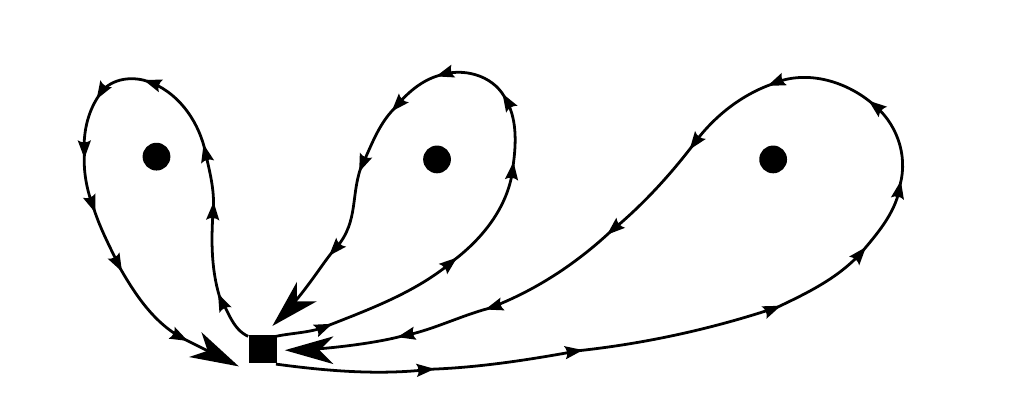
\end{center}
\vskip -0.5cm
\caption{The three different paths around the marked points
  $0,1,\infty$ starting in the lower half plane.
}
\label{fig:paths_in_H-}
\end{figure}

We consider the loops\footnote{
Here and in the following we denote the singular point by a subscript in
square parenthesis, e.g. $\gamma_\hyppar{1}$ 
this is to avoid confusion with the
index associated with the brane. Indices associated with the branes
are put in round parenthesis, e.g. $f_{(1)}$.
Moreover we use $\hypbas{1}$ as subscript to denote well adapted
objects for the singular point $z=1$, see for example
eq. (\ref{Barnes-basis-0}). 
} $\gamma^{(+)}_\hyppar{0}$, $\gamma^{(+)}_\hyppar{1}$,
$\gamma^{(+)}_\hyppar{\infty}$ having a base point in the upper half plane
$H^+\equiv H$ \footnote{We define $H=\{z\in\C; \Im\, z\ge 0\}$ and 
$H^-=\{z\in\C; \Im\, z\le 0\}$} and
looping in counterclockwise direction around the marked points 
$z_0=0,1,\infty$
respectively as shown in figure \ref{fig:paths_in_H+}.
We consider also the corresponding loops $\gamma^{(-)}_\hyppar{0}$, 
$\gamma^{(-)}_\hyppar{1}$,
$\gamma^{(-)}_\hyppar{\infty}$ with base point in lower half-plane $H^-$ as
shown in in figure \ref{fig:paths_in_H-}.

More explicitly we define
\begin{equation}
\gamma^{(\pm)}: t\in[0,1] \rightarrow \C,
~~~~\gamma^{(\pm)}(0)=\gamma^{(\pm)}(1)= z_0\in H^{\pm}
.
\end{equation}
We denote $\gamma_\hyppar{a}*\gamma_\hyppar{b}$ 
the loop formed by first going around
$\gamma_\hyppar{a}$ and then $\gamma_\hyppar{b}$, i.e
\begin{equation}
\gamma_\hyppar{a}*\gamma_\hyppar{b}(t)=
\left\{
\begin{array}{cc}
\gamma_\hyppar{a}(2t) & t\in[0, 1/2]
\\
\gamma_\hyppar{b}(2t-1) & t\in[1/2, 1]
\end{array}
\right.
.
\end{equation}
e have then that 
\begin{equation}
\gamma_\hyppar{0}^{(+)}*\gamma^{(+)}_\hyppar{1}*\gamma^{(+)}_\hyppar{\infty}=1,
~~~~
\gamma_\hyppar{\infty}^{(-)}*\gamma^{(-)}_\hyppar{1}*\gamma^{(-)}_\hyppar{0}=1
.
\label{trivial_paths_products}
\end{equation}
These equations can be generalized to $N$ points with coordinates $z_i$ such that $|z_{i}| < |z_{i-i}|$ as
\begin{equation}
\gamma_\hyppar{z_N}^{+}*\gamma^{+}_\hyppar{z_{N-1}}
*\gamma^{+}_\hyppar{z_{N-2}}\dots*\gamma^{+}_\hyppar{z_1}=1,
~~~~
\gamma_\hyppar{z_1}^{-}*\gamma^{-}_\hyppar{z_{2}}
*\gamma^{-}_\hyppar{z_{3}}\dots*\gamma^{+}_\hyppar{z_N}=1,
.
\label{gen_trivial_paths_products}
\end{equation}

\begin{figure}[hbt]
\begin{center}
\def\svgwidth{250px}
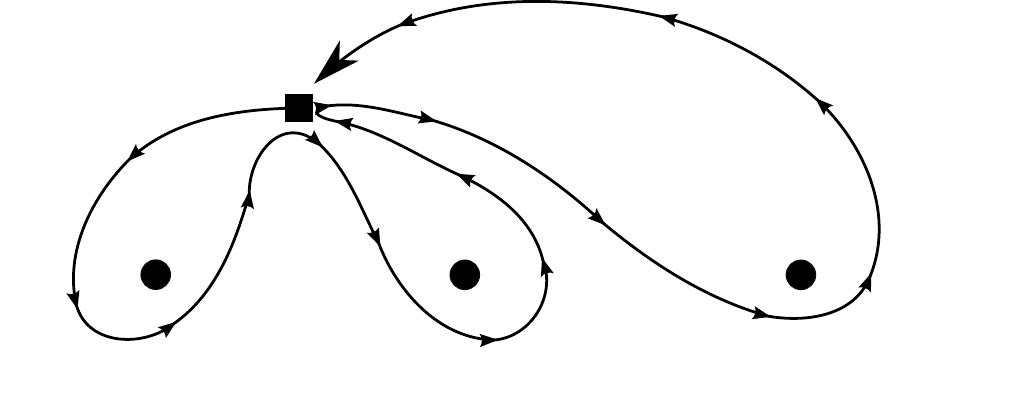
\end{center}
\vskip -0.5cm
\caption{The product 
$\gamma_\hyppar{0}^{(+)}*\gamma^{(+)}_\hyppar{1}*\gamma^{(+)}_\hyppar{\infty}=1$.
}
\label{fig:g0_g1_ginf_Hplus}
\end{figure}

\begin{figure}[hbt]
\begin{center}
\def\svgwidth{250px}
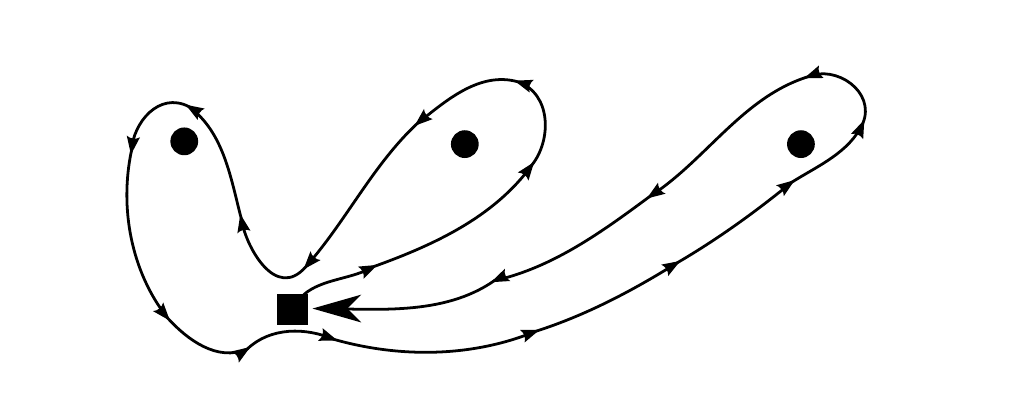
\end{center}
\vskip -0.5cm
\caption{The product 
$\gamma_\hyppar{1}^{(-)}*\gamma^{(-)}_\hyppar{0}*\gamma^{(-)}_\hyppar{\infty}=1$.
}
\label{fig:g1_g0_ginf_Hminus}
\end{figure}

The first equation in eq. (\ref{trivial_paths_products}) is shown in figure
\ref{fig:g0_g1_ginf_Hplus} and and equivalent version of 
the second one, i.e.
$\gamma^{(-)}_\hyppar{1}*\gamma^{(-)}_\hyppar{0}*\gamma_\hyppar{\infty}^{(-)}=1 $
 in figure
\ref{fig:g1_g0_ginf_Hminus}.

The reason why we find two different products in the upper and lower
half plane is simple.
Not all the paths $\gamma^{(+)}$ do transform into
the $\gamma^{(-)}$ ones when we move the base point from the upper half
plane to the lower one.
Moreover there are three different ways of moving a point from the upper
half plane to the lower one. 
These ways are characterized by the two marked points between which
we move the base point.
There are therefore three different possibilities.

\begin{figure}[hbt]
\begin{center}
\def\svgwidth{250px}
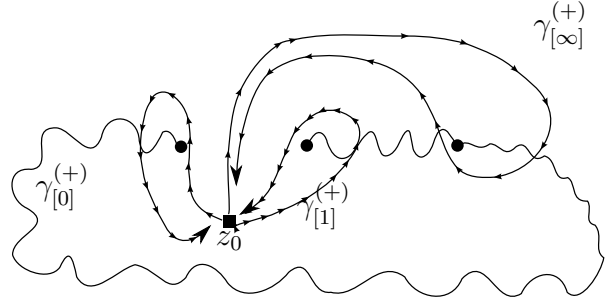
\end{center}
\vskip -0.5cm
\caption{Moving the base point between $0$ and $1$ does not map 
$\gamma^{(+)}_\hyppar{\infty}$ into $\gamma^{(.)}_\hyppar{\infty}$.
}
\label{fig:Change_monodromy_inf}
\end{figure}

\begin{figure}[hbt]
\begin{center}
\def\svgwidth{250px}
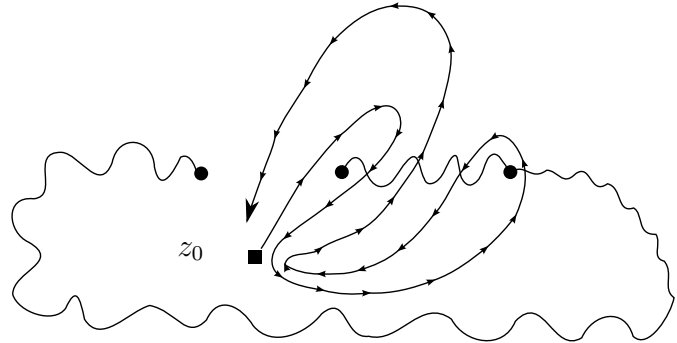
\end{center}
\vskip -0.5cm
\caption{$\gamma^{(+)}_\hyppar{\infty}$ is transformed into 
$\gamma^{(-)\,-1}_\hyppar{1}*\gamma^{(-)}_\hyppar{\infty}*\gamma^{(-)}_\hyppar{1}$.
}
\label{fig:Connect_two_monodromies_inf}
\end{figure}

In figure  \ref{fig:Change_monodromy_inf} we show what
happens when we move the base point from the upper half plane to the
lower half plane between $0$ and $1$.
Both $\gamma^{(+)}_\hyppar{0}$ and $\gamma^{(+)}_\hyppar{1}$ are transformed into the
corresponding paths $ \gamma^{(-)}_\hyppar{0}$ and $\gamma^{(-)}_\hyppar{1}$ while
$\gamma^{(+)}_\hyppar{\infty}$ is transformed into 
$\gamma^{(-)\,-1}_\hyppar{1}*\gamma^{(-)}_\hyppar{\infty}*\gamma^{(-)}_\hyppar{1}$ 
as it is shown in
figure \ref{fig:Connect_two_monodromies_inf}
or equivalently to
$\gamma^{(-)}_\hyppar{0}*\gamma^{(-)}_\hyppar{\infty}*\gamma^{(-)\,-1}_\hyppar{0}$
when we consider the sphere and we move the path on the sphere.
Both these expressions are compatible with eq.s (\ref{trivial_paths_products}).

\begin{figure}[hbt]
\begin{center}
\def\svgwidth{250px}
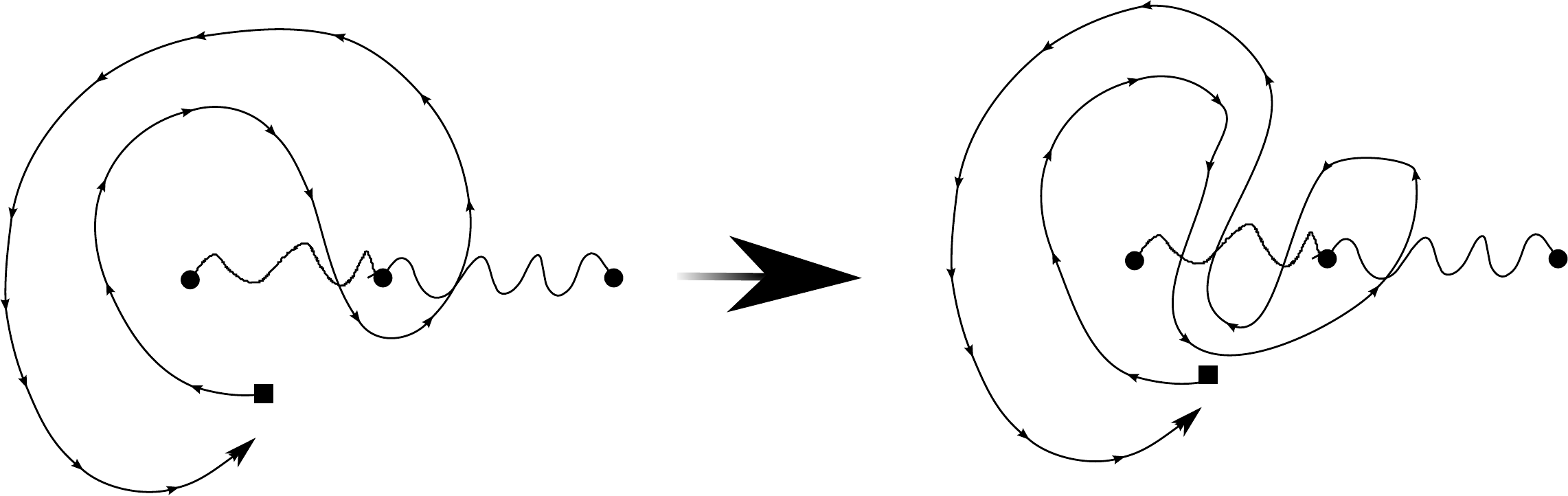
\end{center}
\vskip -0.5cm
\caption{$\gamma^{(+)}_\hyppar{1}$ is transformed into 
$
\gamma^{(-)\,-1}_\hyppar{0}*\gamma^{(-)}_\hyppar{1}*\gamma^{(-)}_\hyppar{0}
\equiv
\gamma^{(-)}_\hyppar{\infty}*\gamma^{(-)}_\hyppar{1}*\gamma^{(-)\, -1}_\hyppar{\infty}
$ .
}
\label{fig:Connect_two_monodromies_1}
\end{figure}

If we move the base point between $0$ and $\infty$ then  
$\gamma^{(+)}_\hyppar{1}$ is transformed into 
$\gamma^{(-)\,-1}_\hyppar{0}*\gamma^{(-)}_\hyppar{1}*\gamma^{(-)}_\hyppar{0}$ when we move
counterclockwise  around the cuts
as shown in figure \ref{fig:Connect_two_monodromies_1}
 and 
$\gamma^{(-)}_\hyppar{\infty}*\gamma^{(-)}_\hyppar{1}*\gamma^{(-)\,
   -1}_\hyppar{\infty}$ when we move clockwise.
The two expressions are nevertheless equal because of the second
equation in (\ref{trivial_paths_products}).

\begin{figure}[hbt]
\begin{center}
\def\svgwidth{250px}
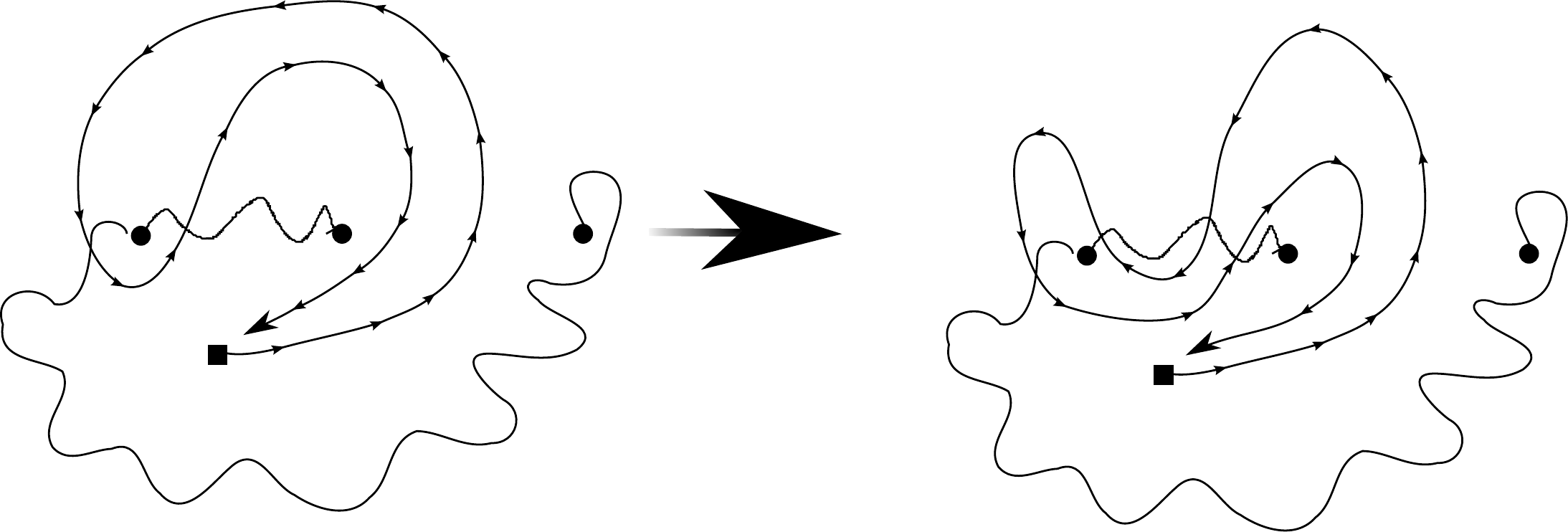
\end{center}
\vskip -0.5cm
\caption{$\gamma^{(+)}_\hyppar{0}$ is transformed into 
$
\gamma^{(-)}_\hyppar{1}*\gamma^{(-)}_\hyppar{0}*\gamma^{(-)\, -1}_\hyppar{1}
$ .
}
\label{fig:Connect_two_monodromies_0}
\end{figure}

Finally in figure \ref{fig:Connect_two_monodromies_0} we show what
happens when we move the base point between $1$ and $\infty$.

\subsection{Hypergeometric equation and its solutions}
As discussed before in order to find a basis of solutions with $U(2)$
monodromies
we start considering the most general Fuchsian differential equation
of order $n=2$ with $N=3$ singularities.
Specializing what described in app. \ref{app:fuchsian} we can write it
as
%
\begin{align}
&\frac{d^2y}{d z^2} 
+ \left[
\frac{1 -\rho_{1 1} -\rho_{1 2} }{z-z_1} +
\frac{1 -\rho_{2 1} -\rho_{2 2} }{z-z_2} +
\frac{1 -\rho_{3 1} -\rho_{3 2} }{z-z_3}
\right]
\frac{d y}{d z}
\nonumber\\
+&\left[
\frac{\rho_{1 1} \rho_{1 2} (z_1-z_2)(z_1-z_3)} {z-z_1}
+\frac{\rho_{2 1} \rho_{2 2} (z_2-z_1)(z_2-z_3)} {z-z_2}
+\frac{\rho_{3 1} \rho_{3 2} (z_3-z_1)(z_3-z_1)} {z-z_3}
\right]
\frac{y}{\prod_{i=1}^{N=3}(z-z_i)}
=0,
\end{align}
where $\rho_{i\, a}$ ($i=1,2,N=3$, $a=1,n=2$) are called the indices
and give the possible behaviors of the solutions around the singular
points, i.e generically we have a mixture as
 $y\sim c_1 (z-z_i)^{\rho_{i\, 1}} + c_2 (z-z_i)^{\rho_{i\, a2}}$.
The indices are constrained as $\sum_{a=1}^{n=2} \sum_{i=1}^{N=3} \rho_{i\, a} =1$.
Its general solution can be formally written as by using the
Papperitz-Riemann $P$-symbol 
\begin{align}
y=
P\left\{ \begin{array}{c c c c  c} 
z_1 & z_2 & z_3 &\\   
\rho_{1\, 1} & \rho_{2\, 1} & \rho_{3\, 1} &z\\   
\rho_{1\, 2} & \rho_{2\, 2} & \rho_{N\, n} & \\   
\end{array} \right\}
.
\label{P-symbol-N3n3}
\end{align}
This symbol represents all the $\infty^2$ solutions of the Fuchsian
equation obtained by the linear combination of two independent
solutions.

Since it represents all the solutions and not one particular solution
it has a number of remarkable properties:
\begin{itemize}
\item 
it is invariant under conformal transformations
\begin{align}
y=
P\left\{ \begin{array}{c c c c  c} 
z'_1 & z'_2 & z'_3 &\\   
\rho_{1\, 1} & \rho_{2\, 1} & \rho_{3\, 1} & z'\\   
\rho_{1\, 2} & \rho_{2\, 2} & \rho_{N\, n} & \\   
\end{array} \right\}
,
\end{align}
with $z'= (a z + b) / (c z + d)$ and $a d - b c =1 $;
\item
it is invariant under columns and lines permutations, for example
\begin{align}
y=
P\left\{ \begin{array}{c c c c  c} 
z'_1 & z'_2 & z'_3 &\\   
\rho_{1\, 2} & \rho_{2\, 1} & \rho_{3\, 1} & z'\\   
\rho_{1\, 1} & \rho_{2\, 2} & \rho_{N\, n} & \\   
\end{array} \right\}
,
\label{P-perm-prop}
\end{align}
is one of the $3!\, 2^2$ cases obtained permuting the singular
points and exchanging their indexes wrt a fixed pair;
\item
it transforms as
\begin{align}
y
=
\left(\frac{z_1-z}{z_3-z}\right)^\delta
\left(\frac{z_2-z}{z_3-z}\right)^\epsilon
P\left\{ \begin{array}{c c c c  c} 
z_1 & z_2 & z_3 &\\   
\rho_{1\, 1} -\delta & \rho_{2\, 1} -\epsilon
                    & \rho_{3\, 1} +\delta+\epsilon &z\\   
\rho_{1\, 2} -\delta & \rho_{2\, 2} -\epsilon& \rho_{N\, n} +\delta+\epsilon& \\   
\end{array} \right\}
.
\end{align}
This is interpreted as the statement that for any solution associated
with the $P$-symbol in eq. (\ref{P-symbol-N3n3}) 
there is one solution associated to the  new $P$-symbol which is
equal to the original 
one when multiplied by $\left(\frac{z_1-z}{z_3-z}\right)^\delta
\left(\frac{z_2-z}{z_3-z}\right)^\epsilon$.
\end{itemize}

Using the last property we can limit ourselves to consider the $P$-symbol
\begin{equation}
y=
\PP
{0}{0}{1-c}
{1}{0}{c-a-b}
{\infty}{a}{b}
,
\label{P-symbol-hyper}
\end{equation}
which is associated with  the hypergeometric equation
\begin{equation}
z(1-z)\frac {d^2y}{d z^2} 
+ \left[c-(a+b+1)z \right] \frac {d y}{d z} -
a b y = 0
,
\label{hypergeo-eq}
\end{equation}
which has singular points $z=0,1 $ and $z=\infty$ where the respective
 indices are $0, 1-c$, $0,c-a-b$ and $a, b$.

This equation has the obvious perturbative solution around the $z=0$
singular point given by\footnote{
While we use the same notation used by NIST the normalization differs.
The reason of this choice is to have simpler monodromy matrices as 
eq. (\ref{Barnes-monodromy-inf}) shows.}
\begin{equation}
y
=
\sum_{n=0}^\infty
\frac{ \Gamma(a+n) \Gamma(b+n) } {\Gamma(c+n) n!} z^n
=
\FF{a}{b}{c}{z}
=
\frac{ \Gamma(a) \Gamma(b) } {\Gamma(c) }
\,{}_2F_1(a,b ; c ; z)
.
\label{FF-0}
\end{equation}
Hence  $\FF{a}{b}{c}{z}$ is among the solutions represented by the
$P$-symbol (\ref{P-symbol-hyper}).
Notice also that $\FF{a}{b}{c}{z}$ is a function defined on the whole
complex plane minus the cut and not only for $|z|<1$ where the series
converges. 

The other independent solution around $z=0$ can be found using the
$P$-symbol properties which yield
\begin{align}
y=
\PP
{0}{0}{1-c}
{1}{0}{c-a-b}
{\infty}{a}{b}
=
(-z)^{1-c}
\PP
{0}{0}{c-1}
{1}{0}{c-a-b}
{\infty}{a-c+1}{b-c+1}
\end{align}
which implies that $(-z)^{1-c} \FF{a+1-c}{b+1-c}{2-c}{z}$ is among the
solutions. 
Since its behavior for $z\rightarrow 0$ is different 
 it is independent of $\FF{a}{b}{c}{z}$.

Now in order to derive the monodromies matrices we need to understand
how the natural basis of the solutions at $z=0$ to the hypergeometric
equation behaves away from the singular point $z=0$ and around the
other singular points.
In the fundamental sheet this basis (in which it has a diagonal
monodromy matrix and has a simple power expansion around $\hypbas{0}$)
 is given by
\begin{equation}
\cB_{\hypbas{0}}(z)
=
\vect
{\FF{a}{b}{c}{z}
}{
(-z)^{1-c} \FF{a+1-c}{b+1-c}{2-c}{z}
}
\label{Barnes-basis-0}
.
\end{equation}

Again these are the functions defined over all the
complex plane minus cuts in the fundamental sheet.
The cuts arise from both from the $\FF{a}{b}{c}{z}$ and from the
$(-z)^{1-c}$.
While the cut from $\FF{a}{b}{c}{z}$ is naturally (but non compulsory)
between $1$ and $+\infty$ the cut from $(-z)^{1-c}$ can be set either
from $0$ to $-\infty$ or from $0$ to $+\infty$.
The former situation is the one depicted in figures
\ref{fig:Change_monodromy_inf} and
\ref{fig:Connect_two_monodromies_inf}
while the latter is the one in figure  \ref{fig:Connect_two_monodromies_1}.
Our choice is to set both cuts along the real positive axis, i.e. 
what it is depicted in figure  \ref{fig:Connect_two_monodromies_1}.

Using again the properties of the P-symbol 
we realize that the basis of solutions in $1/z$ 
(which has a diagonal
monodromy matrix and have a simple power expansion around $\hypbas{\infty}$)
is given by
\begin{equation}
\cB_{\hypbas{\infty}}(z)
=
\vect
{
 \left(- \frac{1}{z} \right)^{a}\,\FF{a}{a+1-c}{a+1-b}{ \frac{1}{z} }
}{
 \left(- \frac{1}{z} \right)^{b} \, \FF{b}{b+1-c}{b+1-a}{\frac{1}{z} }
}
\label{Barnes-basis-inf}
.
\end{equation}


The starting point to connect the two basis is the Barnes integral
representation of a solution of the hypergeometric equation given
by
\begin{equation}
\FF{a}{b}{c}{z}
=
\int_\Gamma \frac{d s}{ 2 \pi i}
\frac{ \Gamma(s+a) \Gamma(s+b) \Gamma(-s)
}{
\Gamma(s+c)
}
(-z)^s
\end{equation}
where $a, b, c \in \C$ and 
$\Gamma$ is the path from $-i \infty$ to $+i \infty$ which has
all the $s=-a-1-n$, $s=-b-1-n$ with $n\in \N$ poles to the left 
and $s=n$ ones to the right.

Notice that the integrand up to the factor $(-z)^s$ 
is a function with only isolated singularities
at finite while the $\FF{a}{b}{c}{z}$ has both isolated singularities
and one cut which originates from the logarithm in $(-z)^s= \exp( \log(-z)
s)$. The cut starts apparently at $z=0$ but the expansion
(\ref{FF-0}) shows at it actually starts at $z=1$.

Notice also that $\FF{a}{b}{c}{z}$ is a function defined on the whole
complex plane minus the cut and not only for $|z|<1$.

The presence of the  logarithmic cut means that we must specify the
value of the logarithm in order to compute the integral and establish
its existence. 
Moreover for seeing in a clear way the connection of the previous
function  with the usual hypergeometric ${}_2F_1$ and therefore 
the absence of the cut between $z=0$ and $z=1$ we can expand it
for $|z|<1$.
To do so we close the path $\Gamma$ on the left but this can be done only if
$-\pi < arg(-z) < \pi$ because  the modulus of the integrand behaves
as 
$|z|^{Re(s)} e^{ Im(s)\, [-\pi\, sign(arg(s)-arg(-z)]}$ for large $|s|$.
When we close on the left we find the original perturbative solution
(\ref{FF-0}) 
\begin{equation}
\FF{a}{b}{c}{z}
=
\sum_{n=0}^\infty
\frac{ \Gamma(a+n) \Gamma(b+n) } {\Gamma(c+n) n!} z^n
=
\frac{ \Gamma(a) \Gamma(b) } {\Gamma(c) }
\,{}_2F_1(a,b ; c ; z)
.
\end{equation}
In a similar way for $|z|>1$ and  still with $-\pi < arg(-z) < \pi$ 
we can close the $\Gamma$ path on the right and get for $b-a\not\in\Z$
\begin{align}
\FF{a}{b}{c}{z}
=&
\sum_{n=0}^\infty
\frac{ \Gamma(b-a-n) \Gamma(a+n) } {\Gamma(c-a-n)\, n!} (-1)^n (-z)^{-a-n}
\nonumber\\
+&
\sum_{n=0}^\infty
\frac{ \Gamma(a-b-n) \Gamma(b+n) } {\Gamma(c-b-n)\, n!} (-1)^n
(-z)^{-b-n}
\label{FF-inf}
\\
&=
\frac{ \sin[\pi(c-a)] } {\sin[\pi(b-a)] } \left( -\frac{1}{z} \right)^{-a}
\, \FF{a}{a+1-c}{a+1-b}{ \frac{1}{z} }
\nonumber\\
&-
\frac{ \sin[\pi(c-b)] } {\sin[\pi(b-a)] } \left(-\frac{1}{z} \right)^{-b}
\, \FF{b}{b+1-c}{b+1-a}{\frac{1}{z} }
,
\label{FF-0-FF-inf}
\end{align}
where the second equality is obtained using the perturbative
definition of  $\FF{a }{b }{c }{\frac{1}{z} }$ given in
eq. (\ref{FF-0}) with argument $1/z$.

Once we have decided where the cuts are we can compute the series
expansion of the basis of solutions (\ref{Barnes-basis-0}) around $z=0$
as in eq. (\ref{FF-0}) or around $z=\infty$ as in eq. (\ref{FF-inf})
by expanding the Barnes integral for $|z|>1$ and closing the path to
the left.

We can then relate the basis of solutions (\ref{Barnes-basis-0}) in
$z$ with the basis of solutions in $1/z$ as
\begin{equation}
\cB_{\hypbas{0}}(z)
=
C\, \cB_{\hypbas{\infty}}(z)
=
%
\frac{1}{\sin[\pi(b-a)] }
\mat
{  \sin[\pi(c-a)]   }
{ -\sin[\pi(c-b)] } 
{ -\sin[\pi a] } 
{ \sin[\pi b] } 
\cB_{\hyppar{\infty}}(z)
\label{Barnes-basis-0-inf}
.
\end{equation}

It is then immediate to compute the monodromies for the basis of
solutions in $z$ (\ref{Barnes-basis-0}) as
\begin{align}
&M_{\hypbas{0} \hyppar{0}}
=
\mat
{1}{0}
{0}{e^{-i 2 \pi c}}
\label{Barnes-monodromy-0}
\\
&
\hspace{-3em}
M_{\hypbas{0} \hyppar{\infty}}
=
\frac{i \, e^{i \pi (a+b)} }{ \sin(\pi c) }
\mat
{ -\cos[ \pi(a+b-c) ] +  e^{-i \pi c} \cos[ \pi(a-b) ] }
{ -2 \sin[ \pi(c-a) ] \sin[ \pi(c-b) ] }
{ +2 \sin[ \pi a ] \sin[ \pi b ] }
{ \cos[ \pi(a+b-c) ] -  e^{+i \pi c} \cos[ \pi(a-b) ] }
,
\label{Barnes-monodromy-inf}
\end{align}
where the second expression comes from the obvious monodromy at
$z=\infty$ for  the basis of solutions in $1/z$
(\ref{Barnes-basis-inf}) 
\begin{align}
M_{\hypbas{\infty} \hyppar{\infty}}
&=
\mat
{e^{i 2 \pi a}}{0}
{0}{e^{i 2 \pi b}}
\label{Barnes-monodromy-inf-inf}
\end{align}
along with eq. (\ref{Barnes-basis-0-inf}).

These monodromy matrices are the same whether we start in the
upper of lower half-plane because of our choice of cuts.
This is not however true  for the monodromy matrices 
$M_{\hypbas{0}\hyppar{1}}^{(\pm)}$ 
around $z=1$
which do generically\footnote{It can happen that the monodromy group
  is abelian and hence there is no difference.} depend on the base point.
From the fact that  monodromy matrices are a 
representation of the homotopy group (\ref{trivial_paths_products}) 
they can be derived  as
\begin{align}
M_{\hypbas{0} \hyppar{1}}^{(+)} 
= M_{\hypbas{0} \hyppar{0}}^{-1}\, M_{\hypbas{0} \hyppar{\infty}}^{-1}
,~~~~ 
M_{\hypbas{0} \hyppar{1}}^{(-)} 
= M_{\hypbas{0} \hyppar{\infty}}^{-1}\, M_{\hypbas{0} \hyppar{0}}^{-1}
\label{z=1_monodromies}
.
\end{align}
A naive way of understanding why this happens is to look at figure
\ref{fig:Connect_two_monodromies_1} and realize that the neighborhood
of $z=1$ is cut into two disconnected pieces by the cuts.

Obviously the same relations hold also for the matrices
$M_{\hypbas{\infty}}$ which are obtained starting from the good basis
at $z=\infty$ (\ref{Barnes-basis-0-inf}) since the two sets of
matrices are connected by a conjugation by the $C$ matrix.

\subsection{$U(2)$ monodromies: constraints on the Papperitz equation}
As sketched in the introduction and better explained in the following
sections we want to use the doublet of solutions 
as the key element to build 
solutions to the string e.o.m with monodromies in $U(2)$.
Therefore we are interested in finding a doublet of functions whose
monodromies belong to $U(2)$.
As it is clear from eq.s (\ref{Barnes-monodromy-0}) 
that $M_{\hypbas{0}\hyppar{0}}$ is in $U(2)$  for real $c$,
$M_{\hypbas{0}\hyppar{\infty}}$ does not generically belong to
$U(2)$.
Therefore the doublet of
solutions of the hypergeometric equation (\ref{hypergeo-eq}) given by the basis
(\ref{Barnes-basis-0}) is not what we are looking for but it is a
close relative.

To fix the problem around $z=0$ and $z=\infty$ we can consider
\begin{align}
\cE_{\hypbas{0}}(z)
&=
(-z)^d \, (1-z)^{f-d}
D_{\hypbas{0}}^{-1}~\cB_{\hypbas{0}}(z)
\nonumber\\
&=
\vect
{ \dzero{1}^{-1} (-z)^d \, (1-z)^{f-d} \FF{a}{b}{c}{z}
}{
\dzero{2}^{-1} (-z)^{d+1-c} \, (1-z)^{f-d} \FF{a+1-c}{b+1-c}{2-c}{z}
}
\label{Basis0}
\end{align}
which is a solution of a Fuchsian equation with three
singular points at $z=0,1$ and $\infty$ and where we have allowed 
for arbitrary complex rescaling  
\begin{equation}
D_{\hypbas{0}}= \mat{\dzero{1}}{}{}{\dzero{2}}
.
\end{equation}

Comparing the indices we see therefore that our solution is within the general
solution represented by the Papperitz-Riemann symbol as
\begin{equation}
\PP
{0}{d}{1-c+d}
{1}{f-d}{c+f-a-b-d}
{\infty}{a-f}{b-f}
\label{Papperitz-Riemann-U2}
.
\end{equation}
Notice that given the previous symbol we can easily find two
independent solutions but generically these solutions will not
generate the desired monodromies. In order to get the desired we need
to normalize  and recombine the solutions to get the solution in
eq. (\ref{Basis0}) back.

We parametrize a $U(2)$ matrix as
\COMMENTO{Check how the change of the $N$ definition 
range changes the mapping $U(2)$ params hypergeo ones.}
\begin{align}
U&= 
\exp(i\, 2\pi\, N ) \exp(i\, 2 \pi\, n^i \sigma_i)
\nonumber\\
&=e^{i\, 2\pi\, N}
\left( \cos(2\pi n) \uno_2 + i \sin(2\pi n) \frac{\vec n}{n}
\cdot \vec \sigma\right),
\end{align}
with
\begin{align}
(N,\vec n)\in
\left\{-\frac{1}{4}\le N < \frac{1}{4},  0\le n \le \oh \right\}/ \sim
~~~~ 
(N,\vec n) \sim (N',\vec n') \mbox{ iff }
N=N', n=n'=\oh 
,
\label{canonical-U2-param}
\end{align}
where $\vec n= (n^1, n^2, n^3)$  and $n= |\vec n|$.
In appendix \ref{app:useful_U2_formula} we report some useful formula
such as the effect on the parameters $N,\vec n$ given by 
the product of two elements  or  the opposite of an element.

We are then interested in 
the relation among the parameters $a, b, c, d, f$,
$\dzero{1}$ and $\dzero{2}$  and the
parameters $N_{\hyppar{0}}, \vec n_{\hyppar{0}}$ and $N_{\hyppar{\infty}}, \vec
n_{\hyppar{\infty}}$ which parametrize the $U(2)$ monodromy matrices 
$U_{\hyppar{0}}
=U(N_{\hyppar{0}}, \vec n_{\hyppar{0}})
=M_{\hypbas{0} \hyppar{0} }$ around $z=0$
and 
$U_{\hyppar{\infty}}=
U(N_{\hyppar{\infty}}, \vec n_{\hyppar{\infty}})$ 
(related to $M_{\hypbas{0} \hyppar{\infty} }$
by a rescaling) around $z=\infty$
for the basis around $z=0$ given by $\cE_{\hyppar{0}}(z)$  in eq. (\ref{Basis0}).

As discussed in the appendix \ref{app:U2_monodromies} 
the monodromy around $z=0$ can only be in the maximal torus
of $U(2)$ and  
we get  $U(2)$ monodromies when the parameters are real
\begin{equation}
a, b, c, d, f \in \R
.
\end{equation}
Moreover they must satisfy the constraint
\begin{equation}
- \sin(\pi a) \sin(\pi b) \sin[\pi (a-c)] \sin[\pi (b-c)] >0
\label{-sin-prod-gr-zero}
\end{equation}
which is necessary in order to be able to find a value
for $\frac{\dzero{2}}{\dzero{1}}$.
In fact the ratio of the moduli of parameters 
$\dzero{1}$ and $\dzero{2}$ is
fixed in order to have $U(2)$ monodromies as
\begin{equation}
\left| \frac{\dzero{2}}{\dzero{1}} \right|^2
=
- \frac{ \sin(\pi a) \sin(\pi b)}{ \sin[\pi (a-c)] \sin[\pi (b-c)]}
\label{modulus_d02/d01}
.
\end{equation}
Their relative phase $e^{i 2\pi \delta_{\hyppar{0}} }$ defined by
\begin{equation}
\frac{\dzero{2}}{\dzero{1}}
=\left| \frac{\dzero{2}}{\dzero{1}} \right| 
e^{i 2\pi  \delta_{\hyppar{0}} },
~~~~
-\oh\le  \delta_{\hyppar{0}} < \oh
\end{equation} 
is arbitrary.
Nevertheless it fixes part of the information on the versor
associated to $\vec n_{\hyppar{\infty}}$
as it enters the last of eq.s (\ref{ni3_ni12}).

It is then immediate to see that the monodromy around $z=0$ 
has $U(2)$ parameters
\begin{align}
2 N_{\hyppar{0}} &= 2 d -c + k_{N_{\hyppar{0}}}
\nonumber\\
2 n_{\hyppar{0}}^3 &= c + k_{n_{\hyppar{0}}}
\nonumber\\
n_{\hyppar{0}}^1 &= n_{\hyppar{0}}^2 =0
\label{N0_n03}
\end{align}
where the integers $k_{N_{\hyppar{0}}}\in\Z$ and $k_{n_{\hyppar{0}}}\in\Z$ 
are uniquely fixed by the range
in which $N_{\hyppar{0}}, n_{\hyppar{0} 3}$ can vary and by
\begin{equation}
k_{N_{\hyppar{0}}} \equiv k_{n_{\hyppar{0}}}~~~~mod~2
.
\end{equation}
Explicitly $0\le N_{\hyppar{0}}< \oh$ fixes $k_{N_{\hyppar{0}}}$ then
this fixes the parity of $k_{n_{\hyppar{0}}}$ and   
$-\oh\le n_{\hyppar{0} 3}< \oh$ fixes it completely.

Similarly from the trace of the monodromy around $z=\infty$ we find
that $U(2)$ has parameters
\begin{align}
2 N_{\hyppar{\infty}} &= a + b - 2 f + k_{N_{\hyppar{\infty}}}
\nonumber\\
2 n_{\hyppar{\infty}} &= (-)^{s_{n\hyppar{\infty}}} (a-b) + k_{n_{\hyppar{\infty}}} 
\label{Ni_ni}
\end{align}
where the integers $k_{N_{\hyppar{\infty}}}\in\Z$ and $k_{n_{\hyppar{\infty}}}\in\Z$
and the ``sign'' $s_{n\hyppar{\infty}}\in\{0,1\}$ are again uniquely
fixed  by the range
in which $N_{\hyppar{\infty}}, n_{\hyppar{\infty}}$ can vary
and by
\begin{equation}
k_{N_{\hyppar{\infty}}} \equiv k_{n_{\hyppar{\infty}}}~~~~mod~2
.
\end{equation}
Moreover we have also
\begin{align}
\frac{n^3_{\hyppar{\infty}}}{n_{\hyppar{\infty}}}
=&
 (-)^{s_{n \hyppar{\infty}} +1 }
\frac{
\cos[ \pi(a+b-c) ] - \cos( \pi c ) \cos[\pi(a-b) ]
}{
\sin( \pi c ) \sin[\pi(a-b) ]
}
\nonumber\\
\frac{n^1_{\hyppar{\infty}} + i\, n^2_{\hyppar{\infty}}}
{ 
n_{\hyppar{\infty}} 
}
=&
e^{-i 2 \pi \delta_{\hyppar{0}} }
(-)^{s_{n_\hyppar{\infty}} +1  }\, sign( \sin(\pi a) \sin(\pi b) )
\nonumber\\
&\frac{ 
\sqrt{-4 \sin(\pi a) \sin(\pi b) \sin[\pi (a-c) ] \sin[\pi (b-c)]
  } 
}{
 \sin(\pi c) \sin[\pi (a-b) ]
}
.
\label{ni3_ni12}
\end{align}

\subsection{From $U(2)$ monodromies to parameters of Papperitz
  equation}
\label{sect:from_U2_monod_to_Papp_eq}
The previous equations can be also inverted.
In this way we can find the parameters $a, b, c, d, f$  and 
$\dzero{ 2}/ \dzero{ 1}$ given the $U(2)$ monodromies at $z=0$ and $z=\infty$.
The former must be in the maximal torus of $U(2)$ generated 
by the Cartan subalgebra and is characterized by
$N_{\hyppar{0}}$ and $n^3_{\hyppar{0}}$.
The latter is a generic $U(2)$ matrix and is fixed by giving
$N_{\hyppar{\infty}}$ and $\vec n_{\hyppar{\infty}}$.

With a simple algebra we find for $(-1)^{s_{ n \hyppar{\infty}}}=+1${} \footnote{
For $(-1)^{s_{ n\hyppar{\infty}}}=-1$ simply exchange $a$ with $b$.
}
\begin{align}
a
&=
n_{\hyppar{0} 3} + n_{\hyppar{\infty}} + (-)^{s_A} A + k_a
\nonumber\\
b
&=
n_{\hyppar{0} 3} - n_{\hyppar{\infty}} + (-)^{s_A} A + k_b
\nonumber\\
c
&=
2 n_{\hyppar{0} 3} + k_c
\nonumber\\
d
&=
n_{\hyppar{0} 3} + N_{\hyppar{0}} + k_d
\nonumber\\
f
&=
n_{\hyppar{0} 3} - N_{\hyppar{\infty}} + (-)^{s_A} A + k_f
,
\label{u2-to-abcdf}
\end{align}
where all $k$s are arbitrary integers, $(-)^{s_A}\in\{\pm1\}$ 
and the quantity $A$ with 
$0\le A < 1/2$ is defined as
\footnote{
Because of the relation eq. (\ref{z=1_monodromies}) among the monodromies
we can expect that $A$ is connected with $n_{\hyppar{1}}$.
Using eq. (\ref{U2U2_product}) we can establish in a more precise way
this relation.

When  
$-\frac{1}{4}\le -N_{\hyppar{0}}-N{\hyppar{\infty}}< \frac{1}{4}$,
i.e.  $-N_{\hyppar{0}}-N{\hyppar{\infty}}$ is in the proper definition
range 
the quantity $A$ is actually $n_{\hyppar{1}}$  
the modulus of the vector $\vec n_{\hyppar{1}}$ which parametrizes 
the $SU(2)$ part 
of the upper half plane monodromy matrix $M_{\hyppar{0} \hyppar{1}}^{(+)}$ in 
$z=1$  as defined in  eq. (\ref{z=1_monodromies}).
Moreover $N_{\hyppar{1}}=-N_{\hyppar{0}}-N{\hyppar{\infty}}$.

When  
$-\frac{1}{2}< -N_{\hyppar{0}}-N{\hyppar{\infty}}< -\frac{1}{4}$,
the quantity $A$ is actually $\oh - n_{\hyppar{1}}$  
and  $N_{\hyppar{1}}=\oh -N_{\hyppar{0}}-N{\hyppar{\infty}}$.

Finally, when  
$\frac{1}{4}\le -N_{\hyppar{0}}-N{\hyppar{\infty}}\le \frac{1}{2}$,
the quantity $A$ is actually $\oh - n_{\hyppar{1}}$  
and  $N_{\hyppar{1}}=-\oh -N_{\hyppar{0}}-N{\hyppar{\infty}}$.

}
\begin{align}
\cos(2 \pi A) 
=
\cos(2\pi n_{\hyppar{\infty}} ) \cos(2\pi n_{\hyppar{0} 3} ) 
- \sin(2\pi n_{\hyppar{\infty}} ) \sin(2\pi n_{\hyppar{0} 3} ) \frac{ n_{\hyppar{\infty} 3} }{ n_{\hyppar{\infty}}} 
.
\label{A-definition}
\end{align}
In order to fix almost completely the solution we need also
\begin{align}
\left | \frac{\dzero{ 2 }}{\dzero{ 1}} \right|^2
&=
-\frac{
\sin(\pi a) \sin(\pi b)
}{
\sin[\pi (a-c) ] \sin[\pi (b-c) ]
}
\nonumber\\
\frac{\dzero{ 2 }}{\dzero{ 1}} / 
\left | \frac{\dzero{ 2 }}{\dzero{ 1}}\right|
&=
e^{-i 2 \pi    \delta_{\hypbas{0}} }
\nonumber\\
&
=
-\frac{
n_{\hyppar{\infty} 1} - i\, n_{\hyppar{\infty} 2}
}{
\sqrt{ n_{\hyppar{\infty} 1}^2 + n_{\hyppar{\infty} 2}^2}
}
\,sign\left(
\sin(\pi a) \sin(\pi b)
\sin(\pi c) \sin[\pi (a-b)]
\right)
.
\label{d0-ratio}
\end{align}
We then fix completely our definition of the  solution by choosing
\begin{align}
D_{\hypbas{0}}
&=
\mat{ 1 }{}
{}{ \left | \frac{\dzero{ 2 }}{\dzero{ 1}} \right| 
e^{-i 2 \pi    \delta_{\hypbas{0}} }}
,~~~~-\oh\le\delta_{\hypbas{0}}<\oh
.
\end{align}
Using the explicit values of the parameters it is possible to verify
that
$$
-{\sin(\pi a) \sin(\pi b)}{\sin[\pi (a-c) ] \sin[\pi (b-c) ]}
=
\sin^2(2\pi n_{\hyppar{\infty}} ) \sin^2(2\pi n_{\hyppar{0} 3})
\left[
1-  \left( \frac{ n_{\hyppar{\infty} 3} }{ n_{\hyppar{\infty}}} \right)^2
\right]
$$
and therefore that the previous expression for 
$\left | \frac{\dzero{ 2 }}{\dzero{ 1}}\right|$
is always  meaningful.

At first sight the presence of the sign ambiguity
$(-)^{s_A}\in\{\pm1\}$ would hint to the existence of two 
different families of solutions where each member of any class 
is labeled by the integers $k$s.
It is not so.
This can be seen using Euler relation
\begin{equation}
 \FF{a}{b}{c}{z}= (1-z)^{c-a-b}  \FF{c-a}{c-b}{c}{z}
.
\end{equation}
Denoting by $\tilde a, \tilde b, \dots$ the quantities for
$(-)^{s_A}=-1$ it is easy to verify that
\begin{align}
(-z)^{\tilde d} \, (1-z)^{\tilde f- \tilde d} \FF{\tilde a}{\tilde b}{\tilde c}{z}
&=
(-z)^d \, (1-z)^{f-d} \FF{a}{b}{c}{z}
\end{align}
when
$\tilde  k_a = k_c -k_b$,
$\tilde  k_b = k_c -k_a$,
$\tilde  k_c = k_c$,
$\tilde  k_d = k_d$ and
$\tilde  k_f = k_f + k_c - k_a - k_b$.
The analogous relation for the second component of $\cE$ is then also
satisfied.

This equivalence can be seen  in a less precise way 
by comparing the Papperitz-Riemann symbols associated to the
two solutions.
Using eq. (\ref{Papperitz-Riemann-U2}) we write for the $(-)^{s_A}=+1$ solution
\begin{equation}
\scriptsize
{\PP
{0}{n_{\hyppar{0} 3} + N_{\hyppar{0}} + k_d}{-n_{\hyppar{0} 3} + N_{\hyppar{0}} +  k_d-k_c+1}
{1}{A -N_{\hyppar{0}}- N_{\hyppar{\infty}} +k_f-k_d}
   {-A -N_{\hyppar{0}}- N_{\hyppar{\infty}} +k_f+k_c-k_a-k_b-k_d}
{\infty}{n_{\hyppar{\infty}}+N_{\hyppar{\infty}}+k_a-k_f}{-n_{\hyppar{\infty}}+N_{\hyppar{\infty}}+k_b-k_f}
}
\end{equation}
and for the $(-)^{s_A}=-1$ solution
\begin{equation}
\scriptsize
\PP
{0}{n_{\hyppar{0} 3} + N_{\hyppar{0}} + \tilde k_d}
   {-n_{\hyppar{0} 3} + N_{\hyppar{0}} +  \tilde k_d-\tilde k_c+1}
{1}{-A -N_{\hyppar{0}}- N_{\hyppar{\infty}}+\tilde k_f-\tilde k_d}
   {+A -N_{\hyppar{0}}- N_{\hyppar{\infty}} +\tilde k_f+\tilde k_c-\tilde k_a-\tilde k_b-\tilde k_d}
{\infty}{n_{\hyppar{\infty}}+N_{\hyppar{\infty}}+\tilde k_a-\tilde k_f}
        {-n_{\hyppar{\infty}}+N_{\hyppar{\infty}}+\tilde k_b-\tilde k_f}
.
\end{equation}
The two symbols coincide  again when we use the previous identifications
 therefore the two families are actually the same.
\COMMENTOO{NON E' LO STESSO??}

As we have discussed above there are actually two possible monodromies
at $z=1$ (\ref{z=1_monodromies}) depending on whether our base point
is in the upper or lower half plane, nevertheless since the $SU(2)$ in
$z=0$ is in the maximal torus, i.e. $\vec n_{\hyppar{0}} =
n_{\hyppar{0} 3} \vec k$ there is no difference in the modulus of $
\vec n_{\hyppar{1}}$.
Actually also $ n_{\hyppar{1} 3}$ is invariant and only $
n_{\hyppar{1} 1}$ and 
$n_{\hyppar{1} 2}$ are different.
%
\COMMENTO{
 because of a transformation property of the Papperitz-Riemann
symbol which for the case of our interest reads
\begin{align}
w=
\PP
{0}{\alpha}{\alpha'}
{1}{\beta}{\beta'}
{\infty}{\gamma}{\gamma'}
=
(1-z)^\delta
\PP
{0}{\alpha}{\alpha'}
{1}{\beta-\delta}{\beta'-\delta}
{\infty}{\gamma+\delta}{\gamma'+\delta}
.
\end{align}
In facts for example the first component of
the doublet $\cE_{\hyppar{0} }$ of the first family, the one with
$(-)^{s_A}=+1$, 
is connect with the second component of the second family, the one
with  $(-)^{s_A}=-1$,  with a different set of $k$s, explicitly
\begin{align}
\cE_{\hyppar{0} 1}&(a, b, c, d, f; (-)^{s_A}=+1; z)=
\nonumber\\
&=\cE_{\hyppar{0} 2}(a-2 A + 2 k_{a b c}, b-2 A + 2 k_{a b c}, c, d, f-2 A +
2 k_{a b c }; (-)^{s_A}=-1; z)
\end{align}
with $a+b-c = 2 A + 2 k_{a b c}= 2 A + k_a + k_b - k_c$.
That $k_a + k_b - k_c$ is even follows from ...
} 

\subsection{The complete abelian solution cannot be recovered}
It is then interesting to take the abelian limit  of the $U(2)$
monodromies,
i.e. consider the case where all monodromies are in the maximal torus $U(1)^2$.
This in order to make contact with the previous papers dealing with
the factorized cases.
Naively this seems to be possible since we have monodromies in $U(2)$
but this is not actually the case.
One intuitive reason is that we are not considering the general
monodromies in $R^4$.
We can nevertheless obviously obtain a $U(1)$ monodromy at the price
of setting to zero one of the two solutions of the $U(2)$ case.

More technically the reason why we cannot obtain solutions with
$U(1)^2 \subset U(2)$ monodromy is the following.
Suppose we want to find a second order equation which has as solutions
the following two abelian solutions 
\begin{align}
w_1&= (-z)^{\epsilon_{\hyppar{0} 1} } (1-z)^{\epsilon_{\hyppar{1} 1} }
\nonumber\\
w_2&= (-z)^{\epsilon_{\hyppar{0} 2} } (1-z)^{\epsilon_{\hyppar{1} 2} }
,
\end{align}
then the indices would be 
$\epsilon_{\hyppar{0} 1}, \epsilon_{\hyppar{1} 1},   
\epsilon_{\hyppar{\infty} 1}= -\epsilon_{\hyppar{0} 1}-\epsilon_{\hyppar{0} 1}$
and
$\epsilon_{\hyppar{0} 2}, \epsilon_{\hyppar{1} 2},   
\epsilon_{\hyppar{\infty} 2}= -\epsilon_{\hyppar{0} 2}-\epsilon_{\hyppar{0} 2}$.
It follows immediately that the sum of the indices is zero and
therefore there cannot exist a Fuchsian second order equation with
these solutions since for any Fuchsian second order equation with
three singularities the sum of indices is one.

This can also be confirmed directly computing the associated second order
differential equation for $w$ as
\begin{equation}
\left|\begin{array}{ccc}
w'' & w' & w \\
w_1'' & w_1' & w_1 \\
w_2'' & w_2' & w_2 
\end{array}
\right|=0
,
\end{equation}
and checking that the leading behavior for $z\rightarrow \infty$ for
the coefficient of $w$ is $O\left(\frac{1}{z^2}\right)$ while it
should be $O\left(\frac{1}{z^4}\right)$ in order to be Fuchsian.

In any case it is interesting to consider how far we can go trying to
recover the abelian solutions.
We can actually recover one of the two abelian solutions while the other
is set to zero because of the scaling coefficients $\dzero{ 1}$ and 
$\dzero{ 2}$.

Explicitly the abelian case corresponds to $\vec n_{\hyppar{0}} =
n_{\hyppar{0} 3} \vec k$ and  
$\vec n_{\hyppar{\infty}} =  n_{\hyppar{\infty} 3} \vec k$.
Then from eq. (\ref{A-definition}) we read immediately that
\begin{equation}
A=
\left\{
\begin{array}{l c}
|n_{\hyppar{0} 3}  + n_{\hyppar{\infty} 3}| & |n_{\hyppar{0} 3}  + n_{\hyppar{\infty} 3}|<\oh 
\\
1-|n_{\hyppar{0} 3}  + n_{\hyppar{\infty} 3}| & |n_{\hyppar{0} 3}  + n_{\hyppar{\infty} 3}|>\oh
\end{array}
\right.
. 
\end{equation}

In any of these cases one of the quantities $a$, $b$, $a-c$ and $b-c$
is an integer and therefore the ratio $\frac{\dzero{ 2 }}{\dzero{ 1}}$
is either zero or infinity.  This means that either $\dzero{ 1}$ or
$\dzero{ 2}$ is zero and hence that one of the components of the
vector of the basis solutions $\cE_{\hyppar{0}}(z)$ is zero.
Another way of explaining this it is to notice that the monodromy
matrix at $z=\infty$ cannot ever become diagonal but at most
triangular.
\COMMENTOO{
Nevertheless as we discuss later the classical string solution has
both a right and left moving parts and each of them contributes an
abelian solution in such a way the a solution for both complex
direction is recovered but the parameters of the two are not independent
as it would be in the general abelian case.
}

\section{String action and branes configuration}
\label{sect:string-action-branes-conf}
Our aim is to describe the configuration of three $D2$ branes
in $R^4$ with global monodromies in $U(2)\subset SO(4)$.
Here we discuss the simplest possible setting.
\COMMENTO
{therefore 
we discuss here the more restrictive setup which leads to some of
those configurations here  and the general setup in the appendix
\ref{app:general_setup_for_U2}. 
}

The part of our interest  of the Euclidean action for the string 
in conformal gauge is given by
\begin{align}
S_E
&=
\frac{1}{2 \pi \alpha'} 
\int_{H} d^2 u ~ \oh \du X^I  \dub X^I 
\nonumber\\
&=
\frac{1}{2 \pi \alpha'} 
\int_{H} d^2 u ~ \oh 
\left( \du Z^i  \dub \bZ^i  + \dub Z^i  \du \bZ^i 
\right)
\label{Classical-action}
\end{align}
where $u,v,\dots \in H$ belong to the upper half plane $H$ 
($\Im\, u \ge 0$), 
$I=1,\dots 2N$ (for the case of our interest $N=2$) 
are the labels of flat coordinates and
$Z^i= \frac{1}{\sqrt{2}} (X^i+ i X^{N+i})$ are complex flat
coordinates with $i=1,\dots N$.
The complex string coordinate is a map from the upper half plane to a
real surface with boundaries in $\R^4$.

\begin{figure}[hbt]
\begin{center}
\def\svgwidth{300px}
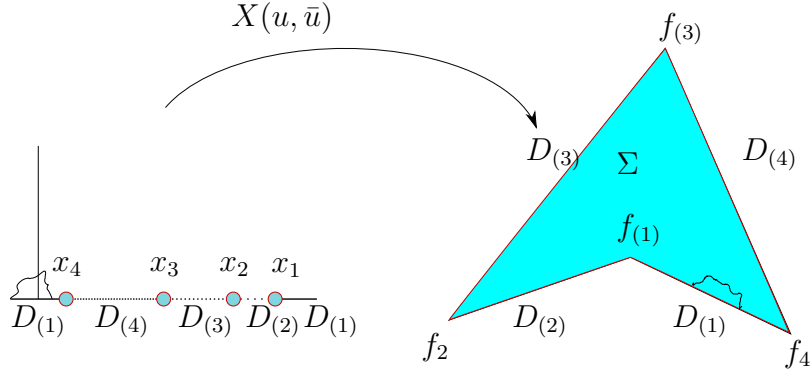
\end{center}
\vskip -0.5cm
\caption{Map from the upper half plane to the target polygon
  $\Sigma$ with untwisted in and out strings.
The map $X(u,\bu)$ folds the boundary of the upper half plane 
starting from $x=-\infty$ in a counterclockwise direction and
preserves the orientation.
}
\label{fig:H2polygon}
\end{figure}

In the cases considered previously the map was  from the upper half plane to a
polygon $\Sigma$ in $\C\equiv \R^2$, i.e. $X:H \rightarrow \Sigma\subset \C$.
For example in fig. \ref{fig:H2polygon} 
we have pictured the interaction
of $\NB=4$ branes at angles $D_{(t)}$ with $t=1,\dots \NB$. The interaction
between brane $D_{(t)}$ and $D_{(t+1)}$ is in $f_{(t)}\in\C$. 
We use the conventions that index $t$ is defined modulo $\NB$
and that 
\begin{equation}
x_{t}<x_{t-1}.
\end{equation}

\begin{figure}[hbt]
\begin{center}
\includegraphics[scale=0.5]{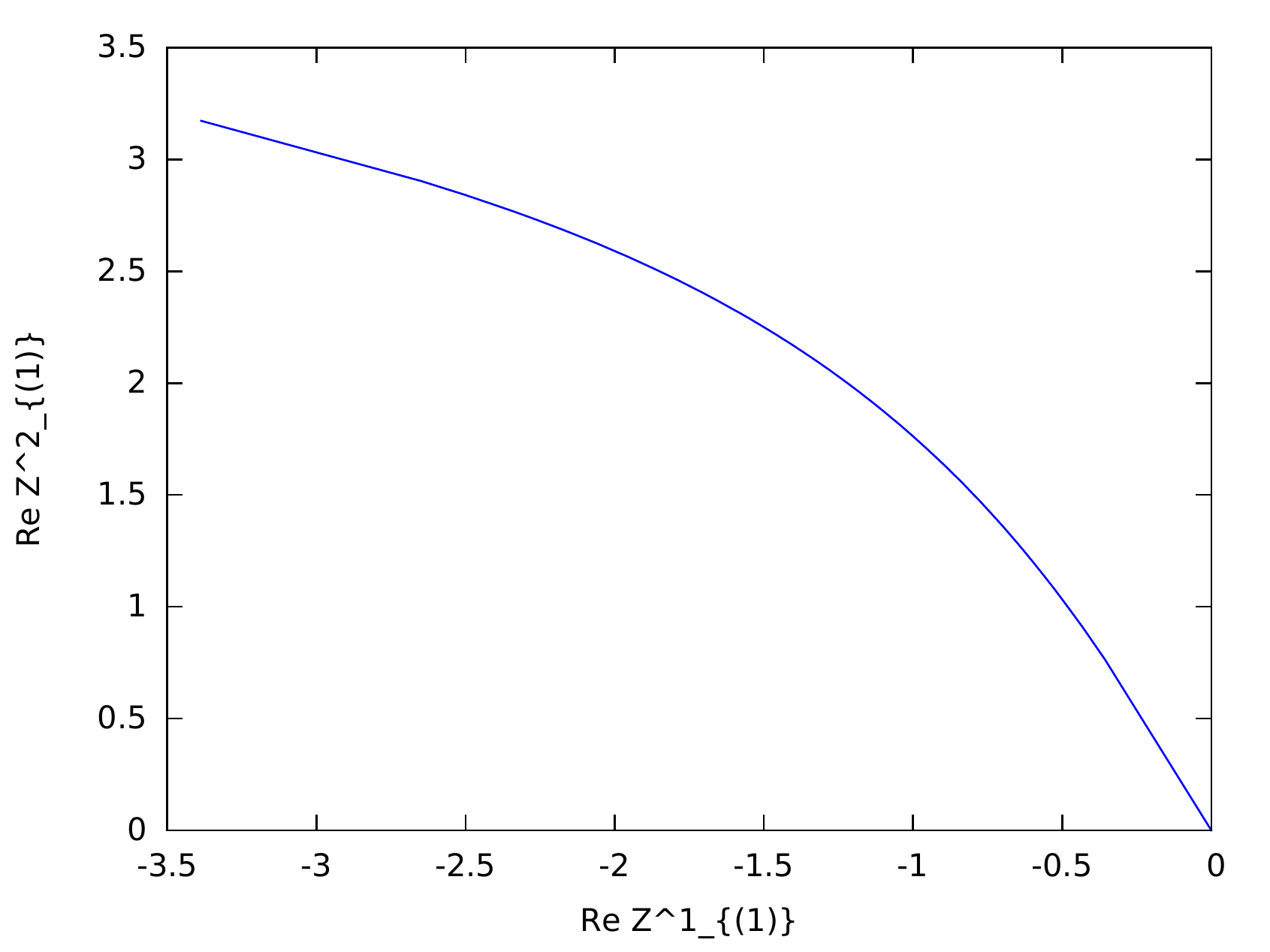}
\end{center}
\caption{
The string endpoint on a brane. A flat surface would intersect a 
brane  along a segment.
Data are with numerical error $<10^{-15}$. 
}
\label{fig:Endpoint}
\end{figure}

In the case we consider we have only three branes and therefore only
three interaction points.
These interaction points always define a 2 dimensional real plane in
$\R^4$ but differently from the cases discussed before in the literature
the embedding of the string worldsheet which follows from the equation
of motion is not  a flat triangle, i.e. a triangle which lies in the plane
determined by the three interactions points.
In fact figure \ref{fig:Endpoint} shows the actual line traced by the endpoint
of the classical string. This line should be compared with 
the naive path which is a segment.

Using the notation described in the next subsection we can describe
more precisely the setup.
The endpoint is shown in good local coordinates for brane
$D2_{(1)}\subset\R^4$ with embedding $\Im Z^i_{(D_1)}=0$.
The embedding matrices are 
\begin{align*}
U_{(1)}= e^{i 2\pi\, 0.4\, \sigma_3},~
U_{(2)}= e^{i 2\pi\, ( 0.3\, \sigma_1+ 0.4\, \sigma_2 + 0.5\, \sigma_3)},~
U_{(3)}= e^{i 2\pi\, ( 0.5\, \sigma_1+ 0.6\, \sigma_2 + 0.7\, \sigma_3)}
\\
M_{(1)}=
\begin{pmatrix}
{0.276341\,i-0.00249408}&{0.914018-0.296982\,i}\\
{ -0.296982\, i-0.914018}&{-0.276341\,i-0.00249408}1
\end{pmatrix}
\end{align*}

\subsection{Local and global branes configuration}
Any brane $D_{(t)}$ $t=1,\dots \NB$ can be described in locally adapted 
real coordinates $X_{ (D_t) }^I$ as
\begin{equation}
X^{N_{(t)}}_{(D_t)}= 0,
\end{equation}
where $N_{(t)}$ runs over the normal directions.
For a $D2$ in $\C^2\equiv \R^4$ we have $I=1,2,3,4$ and the normal directions 
can be taken to be in two classes either $N_{(t)}=3,4$ 
(more generally $N_{(t)}=N+i$  with $i=1\dots N$)
or $N_{(t)}=1,3$ (more generally $N_{(t)}=i, N+i$  with $i=1\dots N/2$).
The former leads to an embedding in locally adapted complex coordinates
as  $Z_{ (D_t) }^1=0$ while the former leads to
\begin{equation}
\Im Z_{ (D_t) }^1= \Im Z_{ (D_t) }^2=0
\label{basic-embeding}
.
\end{equation}
We want to make contact with what usually done for $D1$
embedded into $\C$ where the embedding is described 
$\Im Z_{ (D_t)}^1=0$
therefore we use this former embedding given in eq.s (\ref{basic-embeding})
even if it is apparently less elegant.
Because of this choice the tangent directions index $T$ runs over the
complementary coordinates which in the $D2$ in $R^4$ case are $T=3,4$.

\begin{figure}[hbt]
\begin{center}
\def\svgwidth{250px}
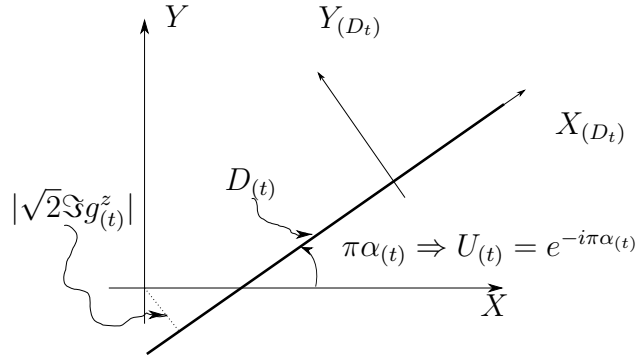
\end{center}
\vskip -0.5cm
\caption{
The relation between local and global coordinates in the simplest case
of $\R^2$. The bold line is the brane $D_t$ with local coordinates 
$Z_{(D_t)}= ( X_{(D_t)}+i Y_{(D_t)} )/ \sqrt{2}$.
It has a distance $|\sqrt{2} \Im g^z_{(t)}|$ from the origin.
}
\label{fig:local_global_coords_inR2}
\end{figure}

The locally adapted complex coordinates are connected to the global
complex coordinates used in defining the string action by a roto-translation
as
\begin{equation}
Z^{i}_{(D_t)} = U^{i}_{(t)\,j} Z^j- g^{i}_{(t)},~~~~
i=1,\dots N
,
\label{Z-loc-Z-glo}
\end{equation}
where the rotation is restricted to $U(N)\subset O(N)$.
This is shown  in figure \ref{fig:local_global_coords_inR2}
in the case of $\R^2$ where we have set
$U_{(t)}= e^{-i \pi \alpha_{t} }$ ($0\le \alpha_{t} < 1$) 
to make contact with the notation
used in previous papers \cite{Pesando:2012cx}.

Our choice is dictated by the fact that we want to find the simplest
configurations which lead to $U(2)$ monodromies of our interest 
\COMMENTO{while
more general configurations can be obtained by choosing the various
rotation matrices $R_{(t)}$ still in $O(N)$ as discussed in appendix
}.  

This means that the local embedding conditions for $D_{(t)}$ can be
written in global coordinates as
\begin{align}
\Im \left[ U^i_{(t)\,j} Z^j- g^i_{(t)} \right]=0,~~~~
i=1,\dots N
,
\label{D2t-embedding}
\end{align}
and hence only $\Im g_{(t)}$ matters for determining the embedding.
Their real d.o.fs  are equal to the number of dimensions transverse to the
brane as it should.

Let us now count the d.o.f..s needed to specify the configuration.
This will teach us something about the kind of configurations we consider.
We start considering the d.o.f..s in rotation matrices only and then we
discuss the full problem with the shifts $\Im g_{(t)}$. 
Would we consider the generic configuration then we would consider the
Grassmannian $ SO(2N)/ SO(N)\times SO(N)$ with $N^2$ real d.o.f..s. On the
other side we do not consider its pure complex version
$ U(N)/ U(N/2)\times U(N/2)$ with $N^2/2$ real d.o.f..s.
To see what we are actually doing we start by rewriting
eq. (\ref{Z-loc-Z-glo}) in real coordinates, explicitly
\begin{align}
X^I_{(D_t)} = R^I_{(t)~J} X^J - g^I_{(t)}
,~~~~
R_{(t)}=\mat
{ \Re U^{i}_{(t)\,j} }{-\Im U^{i}_{(t)\,j} }
{ \Im U^{i}_{(t)\,j} }{ \Re U^{i}_{(t)\,j} }
,~~
g_{(t)}
=\sqrt{2} \vect{ \Re g^i_{(t)} }{ \Im g^i_{(t)} }
,
\label{X-loc-X-glo}
\end{align}
then transformations 
$\vect{X^i_{(D_t)}}{X^{N+i}_{(D_t)}} 
= \mat{O_\parallel}{}{}{O_\perp}
\vect{X^i_{(D_t)}{}' }{X^{N+i}_{(D_t)}{}' }$
with $O_\parallel, O_\perp\in SO(N)$
keep the embedding equations invariant  but destroy the relation
between the good and global coordinates.
This force us to consider $O_\parallel=O_\perp$ hence rotations we
consider are in $ U(N)/ SO(N)$ and have $N (N+1)/2$ real d.o.f..s.
To these d.o.f..s we need then to add other $N$ real dof.s associated
with the shift $\Im g_{(t)}$ in order to completely specify the
embedding.

We conclude therefore that our configuration with $\NB$ branes require
$\NB N (N+3)/2 $ real dof.s to be specified.
This number will be the same when we count it in a different way the
next section.

\subsection{String boundary conditions}
\label{sec:StringBoundaryConditions}
It is immediate to write down the eom associated with the action
(\ref{Classical-action}) as
\begin{equation}
(\partial_x^2 +\partial_y^2) X^I
=\du \dub X^I=0
\end{equation}
along with their general solution as
\begin{equation}
X^I(u,\bu)= 
X^I_L(u)+ X^I_R(\bu) 
,\end{equation}
with $u\in H$.
On this solution we must impose the boundary conditions.
In local real adapted coordinates they read
\begin{equation}
X^{N_{(t)}}_{(D_t)}|_{y=0}
=
\partial_y X^{P_{(t)}}_{(D_t)}|_{y=0}
~~~~x_{t}<x<x_{t-1}
,
\label{bc-real-good-coords}
\end{equation}
where $N_{(t)}$ runs over the normal directions
and $P_{(t)}$ runs over the parallel directions to the brane $D_t$.
The same conditions can be expressed using the well adapted complex
coordinates  as
\begin{equation}
\Im\left[Z^{i}_{(D_t)}\right]|_{y=0}
= \Re\left[\partial_y Z^i_{(D_t)}\right]|_{y=0}=0,
~~~~x_{t}<x<x_{t-1}
,
\label{bc-cpl-good-coords}
\end{equation}
where we have supposed that the string boundary lies on ${D_{(t)}}$ when
$x_{t}<x<x_{t-1}$ 
and that the normal and tangent directions in local coordinates are
always labeled by the same indexes.
These boundary conditions  imply 
(but are not equivalent because of the necessity of taking a
derivative) 
\begin{align}
\vect
{\dub Z_{(D_t) R}(x- i 0^+)}
{\dub \bZ_{(D_t) R}(x- i 0^+)}
 &=
\cR_{(D_t) (Z)}
\vect
{ \du Z_{(D_t) L}(x+i 0^+)}
{ \du \bZ_{(D_t) L}(x+i 0^+)},
\end{align}
where the reflection matrix  $\cR_{(D_t) (Z) (t)}$ 
in local adapted complex coordinates  $Z_{(D_t)}$
 is given by
\begin{equation}
\cR_{(D_t) (Z)}
=
\mat{}{\uno_N}{\uno_N }{}
,\end{equation}
and is idempotent $\cR_{(D_t) (Z)  (t)}^2=\uno_{2 N}$.

In global coordinates the previous equations become
\begin{equation}
\Im\left[ U^i_{(t)\,j}\,  Z^j\right] |_{y=0}- \Im g^i_{(t)}=
\Re\left[ U^i_{(t)\,j}\, \partial_y Z^j\right]|_{y=0}=0,
~~~~x_{t}<x<x_{t-1}
,
\label{Dt-bc}
\end{equation}
Then the global boundary conditions are equivalent to the following conditions 
\begin{align}
\vect
{\dub Z_R(x- i 0^+)}
{\dub \bZ_R(x- i 0^+)}
 &=
\cR_{(Z) (t)}
\vect
{ \du Z_L(x+i 0^+)}
{ \du \bZ_L(x+i 0^+)}
,
~~~~x_{t}<x<x_{t-1},
\nonumber\\
Z(x_t,x_t) &= f_{(t)}
,
\end{align}
where $\cR_{(Z) (t)}$ is an idempotent matrix, 
i.e. $\cR_{(Z)  (t)}^2=\uno_{2 N}$
and $f_{(t)}$ is the intersection point between $D_{(t)}$ and $D_{(t+1)}$.
The matrix $\cR_{(Z) (t)}$ is idempotent since it is
conjugated to  $\cR_{(D_t) (Z)  (t)}$.
Explicitly $\cR_{(Z) (t)}$ is given by
\begin{equation}
\cR_{(Z)\,(t)}
=
\mat
{}{\cU_{(t)}^*} 
{\cU_{(t)}}{} 
=
\mat
{U_{(t)}}{}
{}{U_{(t)}^*}^{-1}
\cR_{(D_t) (Z)}
\mat
{U_{(t)}}{}
{}{U_{(t)}^*}
,
\end{equation}
with 
\begin{equation}
\cU_{(t)}= \cU_{(t)}^T= U_{(t)}^T U_{(t)}.
\label{calU}
\end{equation}
The intersection point\footnote{
It is a point because the codimension of the system of two branes is zero.
} between $D_{(t)}$ and $D_{(t+1)}$ $f_{(t)}$ is
given by
\COMMENTO{ check $2$?}
\begin{align}
f_{(t)}=
2~i~
\left[\cU_{(t)} -\cU_{(t+1)}\right]^{-1}
( U_{(t)}^T  \Im g_t -U_{(t+1)}^T \Im g_{t+1}).
\label{intersection-point-f}
\end{align}

The matrices $\cR_{(Z) (t)}$ are somewhat trivial since they are conjugate
to a reflection matrix and therefore we have
$\cR_{(Z) (t)}=\cR_{(Z) (t)}^{-1}$ and $\det \cR_{(Z) (t)} = (-1)^N$.

Nevertheless we need to specify  $\NB N (N+3)/2$ real dof.s in order to fix the
boundary conditions completely.
$N (N+1)/2$ of these dof.s come from the parameters of the unitary symmetric
matrices $\cU_{(t)}$ ($t=1\dots \NB$) as it can be easily seen by
writing $\cU= \exp(i H)$.
The other $\NB N$ real dof.s  come from the complex vectors $f_{(t)}$
($t=1\dots \NB$). 
These are subject to the following constraints
\begin{align}
\sum_{t=1}^\NB \left[\cU_{(t)} -\cU_{(t+1)}\right] f_{(t)} 
&= 0
,
\nonumber\\
\left[f_{(t)} - f_{(t-1)}\right]^*
&=
\cU_{(t)}
\left[f_{(t)} - f_{(t-1)}\right]
,
\label{f-constraints}
\end{align}
where the first one is actually a consequence of the second set of constraints.
Actually the second set of constraints is simply stating the following
geometrical fact: 
$\{f_{(t)}\}= D_{(t)} \cap D_{(t+1)}$,
$\{f_{(t-1)}\}= D_{(t)} \cap D_{(t-1)}$
and therefore 
$f_{(t)} - f_{(t-1)}\in D_{(t)}$ hence
$\Im \left[ U_{(t)} (f_{(t)} - f_{(t-1)})\right]=0$.

Therefore for our computation of the dof.s
  we need to consider the second set of constraints only  which 
halves the real dof.s in the set $\{f_{(t)} \}$ from $2 \NB N$
real dof.s to $\NB N$.
The meaning of these constraints is roughly to say that we need the
``rotation'' matrices  $\cU_{(t)}$ ($t=1\dots \NB$), 
one corner  $f_{(\bt)}$ for a fixed $\bt$ 
and the ``lengths'' of $\NB-2$ sides to describe the string configuration.

\subsection{String boundary conditions for double fields}
\label{sec:StringBoundaryConditionsForDouble}

Non trivial rotation matrices arise when we use the doubling trick.
In particular we can glue the upper and lower half planes 
along the segment $(x_\bt, x_{\bt-1})$ 
which corresponds to the $\bt$ brane as
\begin{align}
\partial \cZ_{(\bt)}(z)
=
\left\{
\begin{array}{c c}
\du Z_L(u) & z=u\in \mathring{H} \cup  (x_\bt, x_{\bt-1})\\
\cU_{(\bt)}^{-1} \dub \bZ_R(\bu) & z=\bu\in \mathring{H}^-  \cup  (x_\bt, x_{\bt-1})
\end{array}
\right.
,
\nonumber\\
\partial \cbZ_{(\bt)}(z)
=
\left\{
\begin{array}{c c}
\du \bZ_L(u) & z=u\in \mathring{H} \cup  (x_\bt, x_{\bt-1})\\
{\tilde\cU}_{(\bt)}^{-1} \dub  Z_R(\bu) & z=\bu\in \mathring{H}^-  \cup  (x_\bt, x_{\bt-1})
\end{array}
\right.
,
\label{doubling-dZ}
\end{align}
with ${\tilde\cU}_{(\bt)}=\cU_{(\bt)}^{*}$ and
where $\mathring{H} $ is the interior of the upper half plane.

Then the boundary conditions become the discontinuities 
\begin{align}
\dz \cZ_{(\bt)}(x- i 0^+) 
&=  \cU_{(\bt)}^{-1} \cU_{(t)}
\, \dz\cZ_{(\bt)}(x+i 0^+),
\nonumber\\
\dz \cbZ_{(\bt)}(x- i 0^+) &=  
\tilde\cU_{(\bt)}^{-1} \tilde \cU_{(t)}
\, \dz\cbZ_{(\bt)}(x+i 0^+),
~~~~x_{t}<x<x_{t-1},
\end{align}
and the boundary values
\begin{align}
\cZ_{(\bt)}(x_t,x_t) &= f_{(t)}
\label{cZ-global-bc}
.
\end{align}
Notice that the two fields are independent and not connected by a complex 
conjugation even if their monodromies are complex conjugate. 
In fact we find
\begin{align}
\left[ \partial \cZ_{(\bt)}(z) \right]^*&=
\cU_{(\bt)} \partial \cZ_{(\bt)}(w) |_{w=\bz}
\nonumber\\
\left[ \partial \cbZ_{(\bt)}(z) \right]^*&=
\tilde\cU_{(\bt)} \partial \cbZ_{(\bt)}(w) |_{w=\bz}
,
\label{cZ-cbZ-complex-conjugation}
\end{align}
with $\tilde \cU_{(\bt)}=\cU_{(\bt)}^*$.
As discussed in appendix \ref{app:disc2mon} where we pay attention to
the ordering of the generically non commuting matrices
$\cU_{(u)}^{-1} \cU_{(t)}$ 
the previous discontinuities can be rewritten as monodromies as
($\epsilon\in \R$, $0<\epsilon, min( x_{t-1}-x_{t}, x _{t}-x_{t+1})$)
\COMMENTO{DEVO SCAMBIARE upper E lower? NO SOLO le matrici con le inverse: fatto}
\begin{align}
\dz \cZ_{(\bt)}( x_t + e^{i 2 \pi} (\epsilon +i 0^+) ) 
&=
\Mon{t}
  \dz\cZ_{(\bt)}(x_t +\epsilon -i 0^+)
,
~~~~
\Mon{t}
=
\cU_{(t+1)}^{-1} \cU_{(t)}
\nonumber\\
\dz \cbZ_{(\bt)}( x_t + e^{i 2 \pi} (\epsilon +i 0^+) ) 
&=
\widetilde {\Mon{t}}
  \dz\cbZ_{(\bt)}(x_t +\epsilon -i 0^+)
,
~~~~
\widetilde{ \Mon{t} }
=
\Mon{t}^*
\label{dZ-monodromy-in-H}
\end{align}
if we start in the upper half plane and
\begin{align}
\dz \cZ_{(\bt)}( x_t + e^{i 2 \pi} (\epsilon -i 0^+) ) 
&=  
\tMon{t}
\dz\cZ_{(\bt)}(x_t +\epsilon -i 0^+),
~~~~
\tMon{t}
=
 \cU_{(\bt)}^{-1} \cU_{(t)} \cU_{(t+1)}^{-1}  \cU_{(\bt)}
\nonumber\\
\dz \cbZ_{(\bt)}( x_t + e^{i 2 \pi} (\epsilon -i 0^+) ) 
&=  
\widetilde{\tMon{t}}
\dz\cbZ_{(\bt)}(x_t +\epsilon -i 0^+),
~~~~
\widetilde{\tMon{t}}
=
{\tMon{t}}^*
\end{align}
if we start in the lower half plane.
The previous monodromy matrices are not completely
arbitrary $U(N)$ matrices 
since for example $\tMon{t}^T = \cU_{(t+1)}  \tMon{t}^T \cU_{(t+1)}^*$.
Moreover they satisfy a constraint  which follows from the fact 
they are a representation of the homotopy 
group given in eq.s. (\ref{gen_trivial_paths_products}), for example 
in the case $N_B=3$ since $0<x_3<x_2<x_1$ we have
\begin{align}
\Mon{3} \Mon{2} \Mon{1} = \uno
.
\end{align}

While the existence of two sets of monodromies matrices may seem weird 
it is necessary to verify that the Euclidean 
action written using the double fields
\begin{align}
S_E
&=
\frac{1}{8 \pi \alpha'} 
\int_{\C} d^2 z ~ 
\left( 
\dz \cZ_{(\bt)}^i(z)~  \cU^i_{(\bt)\, j} ~\dz \cZ_{(\bt)}^j(w)|_{w=\bz} 
+
\dz \cbZ_{(\bt)}^i(z)~  \cU^{i \,*}_{(\bt)\, j} ~\dz \cbZ_{(\bt)}^j(w)|_{w=\bz}  
\right)
\end{align}
has not any cut.
In the previous expression $\cZ_{(\bt)}^j(w)|_{w=\bz} $ means that 
$\cZ_{(\bt)}^j(w)$ is evaluated for $w\rightarrow\bz$.
Because of this when we go around $x_t$ counterclockwise  
with $z\in H$ we have $\bz\in H^-$  and we go round clockwise in $\bz$
hence the first factor in the first addend contributes 
$[\cU_{(t+1)}^{-1} \cU_{t)}]^T$
and  the second factor in the first addend contributes 
$[ \cU_{(\bt)}^{-1} \cU_{(t)} \cU_{(t+1)}^{-1}  \cU_{(\bt)}]^{-1}$.
Similarly for the tilded fields.

\section{The classical solution}
\label{sect:the_classical_solution}
We are now ready to explicitly compute the classical solution.
We try to be as general as possible as far as possible 
but then we apply the general
procedure to the case of interest, i.e. the $SU(2)$ monodromies.

\subsection{Summary of the previous steps}
Let us summarize the what done up to now and see how we proceed further on.
\begin{itemize}
\item
We start with the embedding data given by $U_{(t)}$ and 
$\Im g_{(t)}$ as follows from the embedding of the $D2_{(t)}$ is given in
eq. (\ref{D2t-embedding}). 
\item
Using these data we can compute the intersection points $f_{(t)}$
between $D_{(t)}$ and $D_{(t+1)}$ as in
eq. (\ref{intersection-point-f}) and the symmetric matrices
$\cU_{(t)}= U_{(t)}^T U_{(t)}$.
\item
We choose a brane $D_{(\bt)}$ which we use for the doubling
trick as in eq.s (\ref{doubling-dZ}).
\item
We can compute the
monodromies matrices $\Mon{t}=\cU_{(t+1)}^{-1} \cU_{(t)}$ as in eq.s 
(\ref{dZ-monodromy-in-H}).
These data are nevertheless independent of the way we perform the
doubling trick and are all what it is needed to compute the
derivatives $\partial\cZ$ and $\partial\cbZ$.
However these matrices do depend on the order of the interaction
points $x_t$ and these can be changed by changing $\Im g_{(t)}$ as it
is easy to see in the simplest abelian case with $\NB=3$. 
\item
The world sheet interaction points are $x_t$ ($t=1,\dots \NB$) where
the twists are inserted. They are singular points which need to be remapped
to the singular points of any possible solution \footnote{
We write any solution and not the solution because there can be many
solutions compatible with all constraints.}.
Moreover the brane $D_{(\bt)}$ for which $x_\bt <x <x_{\bt-1}$ 
where the doubled solution  (\ref{doubling-dZ}) has no cut must be remapped 
to the interval where solutions have no cut.
In analogy with what happens for the hypergeometric function (and for the fixed
singularities of Heun function)
we can assume that the three canonical singular are  $0, 1$ and
$\infty$ and that the interval without cut 
is the real axes interval $(-\infty,0)$ as discussed after eq. (\ref{Barnes-basis-0}).

This remapping can then be done with the $SL(2,\R)$ 
transformation   
\begin{equation}
\omegat(u)= \frac {(u-x_{\bt-1})(x_{\bt+1}-x_{\bt})} 
{(u-x_{\bt})(x_{\bt+1} -x_{\bt-1})}
.
\label{omegabt}
\end{equation}
In particular we have called $\omegat$ the complex variable on which any
solution depends in order not to confuse it with the original doubled
complex variable $z$.

In our case $\NB=3$  and 
the possible solutions are given by any $\cE_{\hypbas{0}}$
compatible with all constraints given in eq. (\ref{u2-to-abcdf}).
If we choose $\bt=1$ then 
the  $SL(2,\R)$ transformation maps $x_3 ,x_2$ and $x^\pm_1$ to $0 ,1$ and
$\mp\infty$ respectively. 
The general mapping is 
\begin{align}
(x_{\bt+1}, x^\pm_\bt, x_{\bt-1}) \rightarrow (1, \pm\infty, 0)
.
\end{align}
\end{itemize}

\subsection{A first naive look}
\label{sect:a_first_naive_look}
After having summarized what done until now we can proceed further
without paying attention to some subtleties which will be addressed in
the next subsection.

\begin{itemize}
\item
In general the monodromy matrix $\Mon{\bt-1}=\cU_{(\bt)}^{-1} \cU_{(\bt-1)}$ at 
$\omegat_{\bt-1}=\omegat(x_{\bt-1})=0$
is not diagonal therefore we need a $U(N)$ ($N=2$ in the case at hand)
transformation $V_{(\bt)}$ to
diagonalize it since we want to use the previous results where the
monodromy matrix $U_{\hyppar{0}}$ at $\omegat=0$ is diagonal.
In particular  we want  
\begin{equation}
U_{\hyppar{0}}
= 
\V{\bt} \Mon{\bt-1} \Vd{\bt}
.
\label{1st-cont-V}
\end{equation}
Naively we could think that given whichever solution of the previous
equation  would work. It is not so as we discuss in the next subsection.
\item
Given the matrix $V_{(\bt)}$ we can map the 
monodromies matrices $\Mon{t}=\cU_{(t+1,t)}$ into the
monodromies matrices of any solution $\cE_{\hypbas{0}}$  given in
eq. (\ref{Basis0}) as
\begin{align}
U_{\hyppar{0}}&= U(N_\hyppar{0}, n_\hyppar{0} \vec k)
= \V{\bt}\, \Mon{\bt-1} \Vd{\bt},~~~~
\nonumber\\
U_{\hyppar{1}}&= U(-N_\hyppar{0}-N_\hyppar{\infty}, \vec n_\hyppar{1} )
= \V{\bt}\, \Mon{\bt+1} \Vd{\bt},~~~~
\nonumber\\
U_{\hyppar{\infty}}&= U(N_\hyppar{\infty}, \vec n_\hyppar{\infty})
=\V{\bt}\, \Mon{\bt} \Vd{\bt}
.
\label{calU-to-hyperU}
\end{align}
Correspondingly the monodromies matrices $\tMon{t}=\cU_{(t,t+1)}^*$ 
for $\dz\cbZ_{(\bt)}$ are mapped into
\begin{align}
{\tilde U}_{\hyppar{0}}
&
= \tV{\bt}\, \tMon{\bt-1} 
\tVd{\bt},~~~~
\nonumber\\
\tilde U_{\hyppar{1}} 
&
= \tV{\bt}\, \tMon{\bt+1} 
\tVd{\bt},~~~~
\nonumber\\
\tilde U_{\hyppar{\infty}}
&
= \tV{\bt}\, \tMon{\bt} 
\tVd{\bt}
.
\end{align}
We can take $\tilde V_{()}= V_{()}^*$ so that 
$\tilde U_{[]}= U_{[]}^*$.
%

When we specialize the previous generic discussion 
to our case with $N=2$ we can parametrize the monodromies as
\begin{equation}
\Ltr{M}{t}{} =U(\Ltr{N}{t}{}, \Ltr{\vec n}{t}{})
.
\end{equation}
Then given $\Ltr{V}{t}{}$ from the previous transformations 
we can compute the vectors $(N_{\hyppar{}},\vec n_{\hyppar{}})$ which are used
to canonically parametrize the matrices $U_{\hyppar{}}$ according to
eq. (\ref{canonical-U2-param}).
Obviously their moduli $n_{\hyppar{}}$ and $N_{\hyppar{}}$ 
are the same of the
corresponding quantities of the monodromies $M_{()}$ since they are
conjugate, i.e we have for example
$n_{\hyppar{0}} = \tilde n_{\hyppar{0}} = n_{(\bt-1)} = \tilde n_{(\bt-1)}$.
In this last expression 
we have also included the relations for the tilded quantities.
\item
Given the parameters $(N_{\hyppar{}},\vec n_{\hyppar{}})$ we can
finally compute the parameters associated to any solution
$\cE_{\hypbas{0}}$, in our case $a,\dots f$ up to integers.
Similarly for the tilded quantities.

Notice that the existence of solutions depends on the possibility of
fixing these integers so that we get a finite Euclidean action.
\item
We can now look for the general solution for the derivative of the
classical string e.o.m (which depend only on the monodromy matrices) 
among the linear combinations 
\begin{align}
\dz\cZ_{(\bt)}(z)
&= 
\frac{\partial \omegat}{ \partial z}
\sum_{r} a_{(\bt) r}
\, \dz\cZ_{(\bt) r}(z)
,&&
\dz\cZ_{(\bt) r}(z)
=
\Vdr{\bt}{r}
\,
 \cE_{\hypbas{0} r}(
\omegat)
,
\nonumber\\
\dz\cbZ_{(\bt)}(z)
&= 
\frac{\partial \omegat}{ \partial z}
\sum_{s} b_{(\bt) s}
\dz\cbZ_{(\bt)z }(z)
,
&&
\dz\cbZ_{(\bt)z }(z)
=
\tVdr{\bt}{s}
\,
 \tilde \cE_{\hypbas{0} s}(
\omegat)
,
\end{align}
where $r$ labels the possible independent solutions  $\cE_{\hypbas{0}  r}$ 
which have the required monodromies  
and finite action as necessary for a classical solution.
We have also allowed for a different $\Vr{\bt}{r}$ for any possible solution.
Because of this we could in principle believe that also the
corresponding monodromies matrices $U_{[]}$ depend on $r$. 
As we show in the next subsection for the case at hand
it is not the case and that the only dependence of
$\Vr{\bt}{r}$ on $r$ is through a phase.
Similarly for $s$ with the tilded quantities.
\item
Finally we can determine the classical solution by fixing the
constants $a$s and $b$s from the global conditions.
In fact we can write the classical solution as
\begin{align}
Z(u,\bu)
=& f_{(\bt-1)}
&&+
\sum_r a_{(\bt) r} \int_{0; \omegat\in H}^{\omegat(u)} d\omegat\, 
V^\dagger_{(\bt) r} \cE_{\hypbas{0} r}(\omegat)
\nonumber\\
&
&&+
\sum_s b_{(\bt) s} \int_{0; \bomegat\in H^-}^{\omegat(\bu)} d\bomegat\, 
{\cU}^*_{(\bt)} {\tilde V}^\dagger_{(\bt) s} 
\tilde \cE_{\hypbas{0} s}(\bomegat)
,
\nonumber\\
\bZ(u,\bu)
=& \bar f_{(\bt-1)}
&&+
\sum_r a_{(\bt) r} \int_{0; \bomegat\in H^-}^{\omegat(\bu)} d\bomegat\, 
{\cU}_{(\bt)}  V^\dagger_{(\bt) r} \cE_{\hypbas{0} r}(\bomegat)
\nonumber\\
&
&&+
\sum_s b_{(\bt) s} \int_{0; \omegat\in H}^{\omegat(u)} d\omegat\, 
{\tilde V}^\dagger_{(\bt) s}  \tilde \cE_{\hypbas{0} s}(\omegat)
,
\label{Z-bZ-general-solution}
\end{align}
and then impose the further global conditions (\ref{cZ-global-bc})
\begin{align}
Z_{(\bt)}(x_t,x_t) &= f_{(t)}
,
~~~~t\ne\bt-1
\end{align}
in order to fix the constants.
\end{itemize}

We can summarize what we have got until now by saying that twist
conditions are local and therefore knowing the modulus of the twists
is enough to determine up to integers the indices.
The sentence ``up to integer'' is important since the sum of all
indices for the Papperitz-Riemann equation, i.e.
the general Fuchsian second order equation with three singularities
must add to $1$ for $\NB=3$  twists.
In fact  we can write the previous statement for the case of
monodromies in $SU(2)$, $\NB=3$ twist fields 
in a sketchy way as
\begin{align}
\cE(\omegat)\sim
\PPom
{0}{n_{(\bt-1)}}{-n_{(\bt-1)}}
{1}{n_{(\bt+1)}}{-n_{(\bt+1)}}
{\infty}{n_{(\bt)}}{-n_{(\bt)}}
\end{align}
where all indices sum to $0$ while they should sum to $1$.
In a more precise way we can write\footnote{
Remember that the $P$ symbol represents all the $\infty^2$ solutions
therefore by writing $=$ it is meant that $y$ is one of these solutions.
}
\begin{align}
\cE(\omegat)=
\PPom
{0}{n_{(\bt-1)} +k_{(\bt-1) +}}{-n_{(\bt-1)} +k_{(\bt-1) -}}
{1}{n_{(\bt+1)} +k_{(\bt+1) +}}{-n_{(\bt+1)} +k_{(\bt+1) -}}
{\infty}{n_{(\bt)} +k_{(\bt) +}}{-n_{(\bt)} +k_{(\bt) -}}
\end{align}
with $\sum_{t=1}^3 \sum_{s\in\{\pm\}} k_{(t) s}=1$.
Now requiring that the action be finite implies that we must set
\begin{align}
\cE(\omegat)=
\PPom
{0}{ -\bn_{(\bt-1)} }{-n_{(\bt-1)}}
{1}{-\bn_{(\bt+1)} }{-n_{(\bt+1)}}
{\infty}{2-\bn_{(\bt)}}{2-n_{(\bt)}}
\end{align}
where we have introduced for notation simplicity 
\begin{equation}
\oh \le \bn=1-n < 1,
\end{equation}
which has however a different range w.r.t. $n$ ($ 0\le n < \oh$).
Similarly we have
\begin{align}
\tilde \cE(\omegat)=
\PPom
{0}{ -\bn_{(\bt-1)} }{-n_{(\bt-1)}}
{1}{-\bn_{(\bt+1)} }{-n_{(\bt+1)}}
{\infty}{2-\bn_{(\bt)}}{2-n_{(\bt)}}
.
\end{align}
Notice that generically the behavior of the solution around any
singular point is given by the sum of the two possible behaviors as for
example in eq. (\ref{FF-0-FF-inf}) which shows that even if we start
from a function having a unique index in a singularity we end up with
a mixture of indices in other singularities. 
This means that we must require the all combinations of the two
indices at any singular point must give a finite action.
For example at $\omegat=\infty$ we require $2(n_{(\bt)} +k_{(\bt) +}) >2 $,
$2(-n_{(\bt)} +k_{(\bt) -}) >2 $ and $(n_{(\bt)} +k_{(\bt) +}) + (-n_{(\bt)}
+k_{(\bt) -}) >2 $. 
In particular the last equation means that $k_{(\bt) +}+ k_{(\bt) -} \ge 4 $.

As it is obvious form the explicit expressions there is an asymmetry 
among points and this is disturbing since the $P$ symbol is invariant
under $SL(2,\C)$ transformations. This asymmetry is however apparent
since the $P$ symbol is directly connected to $\partial_\omegat \cZ$ and not to
$\partial_z \cZ$.
To see how this solve the asymmetry issue consider one of the simplest
cases where we use $\homegat=1/\omegat$ as new variable then  
\begin{align}
\cE(\homegat)=
P\left\{ \begin{array}{c c c c}
{\infty} & 1 & 0 & \\
{-\bn_{(\bt-1)} } & { -\bn_{(\bt+1)} } & { 2 -\bn_{(\bt)} } & \homegat \\
{-n_{(\bt-1)}} & {-n_{(\bt+1)}} & {2-n_{(\bt)}} &
\end{array} \right\}
.
\end{align}
But now $\partial_{\homegat} \cZ = {\homegat}^{-2} \partial_\omegat \cZ$
and hence around $\homegat=0$ we get 
$\partial_{\homegat} \cZ \sim {\homegat}^{-\bn_{(\bt)}} $ and  
$\partial_{\homegat} \cZ \sim {\homegat}^{-n_{(\bt)}} $.
Similarly for $\homegat=\infty$ where we get
$\partial_{\homegat} \cZ \sim {1/\homegat}^{2-\bn_{(\bt)}} $ and  
$\partial_{\homegat} \cZ \sim {1/\homegat}^{2-n_{(\bt)}} $.
This restores completely the symmetry among the points.

Another point on which is worth noticing and commenting is the appearance of an
antiholomorphic part in $Z$ while in the corresponding case with abelian
monodromies this does not happen. 
In fact the previous discussion shows that there is one solution for
$\partial \cZ$ and one for $\partial \cbZ$.
The reason why we get two solutions can be easily understood by
noticing that we have more equations to fix the coefficients $a_r$ and
$b_s$ than in the abelian case.
In fact according to the discussion after eq.s (\ref{f-constraints})
the configuration is determined by $ \NB N (N+3)/2= 15$ real parameters
and the proposed solution (\ref{Z-bZ-general-solution}) depends on
$f_{(\bt-1)} $, $\cU_{(\bt)}$ and $\Mon{t}$ ($t=1,\dots \NB$) 
for a total of $2 N + \NB N(N+1)/2 $ real parameters.
This happens since given $\cU_{(\bt)}$ and $\Mon{t}$ we can compute
all $\cU_{(t)}$ and any symmetric unitary $\cU_{(t)}$ is specified by
$N(N+1)/2$ real dof.s. 
Hence we still need $(\NB-2) N  = 2 $ real equations to fix the
vector  the remaining quantities $f_{(t)}$ ($t\ne\bt-1$), 
in particular we can simply determine $f_{(\bt)}- f_{(\bt-1)}$.
On the other side the reason why we get both a holomorphic and
antiholomorphic contribution to $Z$ is less obvious and may be
traced back to the fact that minimal area surface is not anymore
drawable on a plane despite we have only three interaction points
which uniquely fix a plane in $\R^4$.
This happens because the rotations of the $D2$ branes are not abelian.

This can give the impression that anything can be done easily also for
more complex cases. 
Unfortunately, it is not so. 
Let us see why.

The same approach can be generalized to the $\NB=4$ case with $SU(2)$
global symmetry. So we can write using a
generalized P-symbol in the case $\bt=4$ in a sketchy way
\begin{align}
\cE_{\NB=4, SU(2)}(\Ltr{\omega}{4}{})\sim
P\left\{ \begin{array}{c c c c c c} 
0 & a & 1 & \infty & &\\   
k_{(1)+}-\bn_{(1)} & k_{(2)+}-\bn_{(2)} & k_{(3)+}-\bn_{(3)} &
k_{(4)+}+2-\bn_{(4)} & q 
& \Ltr{\omega}{4}{}\\   
k_{(1)-}-n_{(1)} & k_{(2)-}-n_{(2)} & k_{(3)-}-n_{(3)} & k_{(4)-}+2-n_{(4)} &  &\\   
\end{array} \right\}
,
\label{heun}
\end{align}
where $a$ is the location of the fourth singularity whence we have
fixed the other three and
with $\sum_{t=1}^4 \sum_{s\in\{\pm\}} k_{(t) s} =S =2$
and $k_{(t) s}\ge 0$ in order to have a finite action.
Therefore we have 
$\frac{[S+(2\NB-1)]!}{S! (2\NB-1)!}=\frac{(2+7)!}{2!  7!}=36$ 
possible solutions $\cE$
and $36$ possible $\tilde \cE$s even if we expect to need $(\NB -2) N  =
4 $ solutions only.

The question could be solved by a direct computation if it were not
because of two issues.
The first one is that the general solution of the general 
Fuchsian second order equation with four singularities
 is not uniquely determined by the indices as it happens for the hypergeometric
function but it has an accessory parameter $q$.
In fact, as reviewed in appendix \ref{app:fuchsian}, a Fuchsian equation of
order $N$ with $\NB$ singular points  has  $\NB N(N-1)/2 - N^2 +1$ free
accessory parameters. 
The fixing of the accessory parameters 
is therefore the first issue it is necessary to solve if we want to
consider more complex cases than the actual one.

Another issue and more fundamental is that in order to write the actual
solution we need to normalize the solutions in order to get the
desired monodromies. This is what done in eq. (\ref{Basis0}) for the
case $\NB=3$ and $N=2$ which discuss in this paper.
However as it is clear from the discussion in the present case this
normalization does depend on the continuation formulas for the
different basis of solutions around the different singular points.
Unfortunately this problem is not solved in the general case 
even in the simplest case of Heun function whose $P$ symbol is given
in eq. (\ref{heun})  and 
which corresponds to the second order Fuchsian equation
with four singularities.

At least a couple of possible ways forward can be imagined to try to solve
these issues in this case:
\begin{itemize}
\item
consider special $\Mon{t}$ values which correspond to algebraic
solutions  to the differential equation. This amounts to say that we
are actually working a higher genus Riemann surface as done in the
paper by Inoue \cite{non_abelian};
\item
try to use CFT factorization.
\end{itemize}

\begin{figure}[hbt]
\begin{center}
\def\svgwidth{300px}
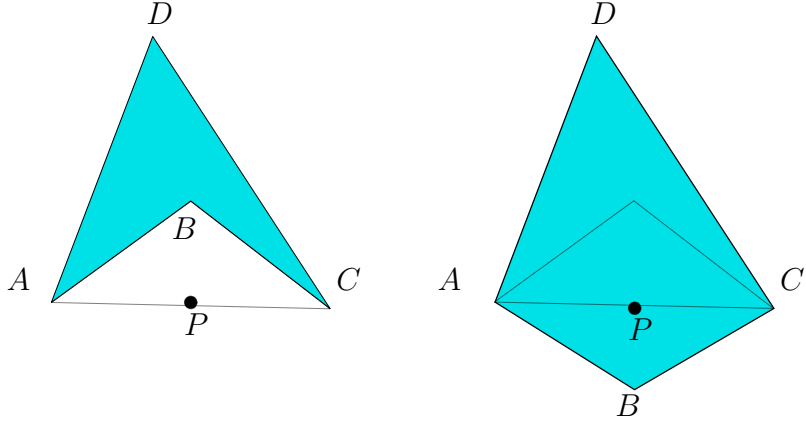
\end{center}
\vskip -0.5cm
\caption{
On the left the case $\NB=4$ $M=1$ and on the right the case $\NB=4$
$M=2$ with the minimal ares shadowed.
}
\label{fig:N4M1_N4M2_in_3dim}
\end{figure}

Another point is worth discussing.
In  the usual factorized case where we consider only $\R^2$ there are
two possible cases for $\NB=4$ as discussed in \cite{Pesando:2012cx}
and they are labeled by an integer $M$ which in this case can be
either $1$ or $2$. The situation is pictured in figure
(\ref{fig:N4M1_N4M2_in_3dim}).
As long as we limit ourselves to the $\R^2$ case we cannot move from
one case to the other by moving the point $B$ without going through
the straight line, i.e. the case of no twist. 
This explains why the two cases are different in $\R^2$.
At first sight this should be not true in $\R^3$ since we can rotate
the curve $A B C$ around the $A P C$ axis 
in a third dimension without going through the straight line. 
Hence we would expect to have only one case.
Actually it is no. 
The reason is that while rotating the curve $A B C$ in the third
dimension in order to deform the left configuration to the right one 
the minimal area bounded by $A B C D$ starts increasing and a certain point  
a second configuration bounded by $A P C D$ and $A B C P$ 
of equal area appears\footnote{
An easy model to see what is going on is to approximate the minimal
area configurations by a sum of triangles.
We consider the simplest case with $A\equiv(0,0,0)$, 
$P\equiv(B,0,0)$, $C\equiv(2B,0,0)$, $D\equiv(B,H,0)$ and 
$B\equiv(B, h \cos \theta , h \sin \theta)$.
This case corresponds to the case where both the triangles $A C D$ and
$A C B$ are isosceles.
Then the approximated area for the $M=1$ case is twice the area of the
triangle $A B D$ and is
$A_{M=1}=\sqrt{H^2 h^2+ B^2 H^2 + B^2 h^2- ( H^2 h^2 \cos^2 \theta + 2
  B^2 \cos \theta) }$
and the approximated area for the $M=2$ case is the sum of the two
triangles $A C D$ and $A B C$ and is
$A_{M=2}= B(H+ h)$.
It is the immediate to see that for the case $B^2 < H h$ the maximum
of $A_{M=1}$  is $\max A_{M=1}= B \sqrt{(H+h)^2+ (H-h)^2}$ which is
greater than $A_{M=2}$.
}.

\COMMENTO{
Two sectors $M=1,2$ for $N=4$ in $U(1)$ case

Use of special $\tMon{t}$ values to use a Riemann surface covering

Use CFT factorization 

Computation of naive possible solutions adding +2 to satisfy indices sum rules
}

The same issues are present also for the $\NB=3$ case with $SU(3)$ global
monodromies where we can expect to write in a sketchy way
\begin{align}
\cE_{\NB=3, SU(3)}(\Ltr{\omega}{3}{})\sim
P\left\{ \begin{array}{c c c c c } 
0 & 1 & \infty & &\\   
n_{(1)} & n_{(2)} & n_{(3)} & q & \Ltr{\omega}{3}{}\\   
m_{(1)} & m_{(2)} & m_{(3)} &  &\\   
-m_{(1)}-n_{(1)} & -m_{(1)}-n_{(2)} & -m_{(3)}-n_{(3)} &  &\\   
\end{array} \right\} 
\end{align}
where $n_{(i)}$ and $m_{(i)}$ are the independent local $SU(3)$ rotation
parameters and again we have one accessory parameter $q$.
Nevertheless we can consider special cases where the monodromies are
known (\cite{beukers}).

\subsection{A more detailed look}
From the discussion of the previous subsection it seems that any
$\V{\bt}$ would do the job since all of them fix the same
indices. However it is not so.

The reason is that there are further constraints from the action of complex
conjugation  on $\dz\cZ_{(\bt)}(z)$ and $\dz\cbZ_{(\bt)}(z)$.
In fact eq.s (\ref{cZ-cbZ-complex-conjugation}) (or eq.s (\ref{Z-bZ-general-solution})) imply for any $r$ and $s$
\begin{align}
\cU_{(\bt)}
&=
\frac{a_{(\bt) r}^*}{a_{(\bt) r}}
\Vr{\bt}{ r}^T\, D^{-1 *}_{\hypbas{0}} D_{\hypbas{0}} \Vr{\bt}{r}
\nonumber\\
\tilde \cU_{(\bt)}
&=
\frac{b_{(\bt) s}^*}{b_{(\bt) s}}
\tVr{\bt}{s}^T\, \tilde D^{-1 *}_{\hypbas{0}} \tilde D_{\hypbas{0}} \tVr{\bt}{s}
,
\label{2nd-cont-V}
\end{align}
since $\cE\sim D^{-1} \cB$ and $\cB$ is a pair of hypergeometric functions
with real parameters. 
These constraints are fundamental to get  a solutions which satisfies
the required boundary conditions, i.e. with the string boundaries on
the branes.

They are however not satisfied by a random solution $\Vr{\bt}{r}$
of eq. (\ref{1st-cont-V}) or the equivalent for $\tVr{\bt}{s}$. 
The problem in implementing them at this stage is that $D_{\hypbas{0}}$
does depend on $\vec n_\hyppar{\infty}$ as in eq.s (\ref{d0-ratio}). 
On the other side  $\vec n_\hyppar{\infty}$ in turn depends 
on $V_{(\bt)}$ because $\vec n_\hyppar{\infty}$ is the parameter
associated with the conjugation of $\cU_{(\bt+1,\bt)}$ by $V_{(\bt)}$
as in the last of eq.s (\ref{calU-to-hyperU}).
Therefore we must solve eq.s (\ref{2nd-cont-V}) and (\ref{calU-to-hyperU})
together.

Let us now solve the previous constraints.
This is done by comparing the previous equations (\ref{2nd-cont-V})
from the behavior of the doubled solution under complex complex conjugation
with the definition $\cU_{(\bt)}= U_{(\bt)}^T U_{(\bt)}$ as given in
eq. (\ref{calU}).
This comparison  suggests to choose the ``gauge''\footnote{
Notice that we can always choose the gauge (\ref{gauge-n2}) by a
``rotation'' in the Cartan group of $SU(2)$ since this does not change 
$U_{\hyppar{0}}$.
}
\begin{equation}
n_{\hyppar{\infty} 2}=0
,
\label{gauge-n2}
\end{equation}
so that we have
\begin{equation}
D^{-1 *}_{\hypbas{0}} D_{\hypbas{0}}=\uno_2
\end{equation}
then we can make the ansatz 
$U_{(\bt)} = e^{-i 2\pi \alpha_{(\bt) r} } \Ltr{R}{\bt}{}\,
\Vr{\bt}{r}$ \footnote{In principle $\Ltr{R}{\bt}{}$ should be written
  as $\Ltr{R}{\bt}{r}$ but as we now show it is actually independent
  on $r$.}
where the phase is defined from
 $a_{(\bt) r}= |a_{(\bt) r}| e^{i 2 \pi \alpha_{(\bt) r} } $
and
\begin{align}
\Ltr{R}{\bt}{}^T\, \Ltr{R}{\bt}{} 
&= 
\uno_2,~~~~
\end{align}
moreover $\Ltr{R}{\bt}{}\in U(2)$ since both $U$ and $V$ are in $U(2)$.
This implies $\Ltr{R}{\bt}{}=U(0,  \Ltr{r}{\bt}{2} \vec j)$ \footnote{
The parametrization given in the main text is such that 
$\Ltr{R}{\bt}{}\in SU(2)$ but we could 
for example have $\Ltr{R}{\bt}{}=U(0, \Ltr{r}{\bt}{2} \vec j)
\sigma_1\in U(2)$.
We can exploit this fact to choose $\vec n_{[0]}= n_{[0]} \vec k$,
i.e. $n_{[0] 3}=n_{[0]}>0$.
}.
Then we get  for any unknown $\Vr{\bt}{r}$
\begin{align}
\Vr{\bt}{r}
&=
 e^{i 2\pi \alpha_{(\bt) r} } \Ltr{R^\dagger}{\bt}{} \, U_{(\bt)} 
.
\label{Vr-ansatz}
\end{align}

We are now left with the problem of computing $\Ltr{R}{\bt}{}$ and
$U_{[\infty]}$ {}%
\footnote{Also $U_{[\infty]}$ turns out to be independent on $r$.}%
.
These are the only unknowns since, as noticed before, $U_{[0]}$ is
completely determined and independent on $r$.
This happens because it is in the Cartan,
i.e. $\vec n_{[0]}= n_{[0]  3} \vec k$ and $M_{(\bt-1)}$ and $U_{[0]}$
are conjugate,
therefore if we write as before
$
M_{(\bt-1)}=U(0, \Ltr{\vec n}{\bt-1}{})
$
 we can always set $n_{[0] 3}= n_{(\bt-1)} >0$.

Because of the same reason $U_{[\infty] r}$ is partially determined and
we must only fix the ratio $n_{[\infty] 1}/n_{[\infty] 3}$.
We have therefore two unknowns $\Ltr{r}{\bt}{2}$ 
and $n_{[\infty] 1}/n_{[\infty] 3}$.
They can be fixed using the first and second equation in the group of  eq.s
(\ref{calU-to-hyperU}).
The first one can be rewritten as
\begin{align}
U^*_{(\bt)} \cU_{(\bt-1)} U_{(\bt)}^\dagger
&=
( U_{(\bt-1)} U_{(\bt)}^\dagger)^T ( U_{(\bt-1)} U_{(\bt)}^\dagger)
=
\Ltr{R}{\bt}{} U_{[0]} \Ltr{R}{\bt}{}^T
.
\end{align}
Since%
\footnote{
Since $U_{[0]}$ is diagonal we can easily compute $U_{[0]}^\oh$    and
then  naively get
$
\Ltr{R}{\bt}{}
= 
\pm ( U_{(\bt-1)} U_{(\bt)}^\dagger)^T  U_{[0]}^\oh
$.
This is however wrong since this would-be solution does not
generically satisfy $\Ltr{R}{\bt}{}^T \Ltr{R}{\bt}{}=\uno$.
}
both the rhs and the lhs are symmetric matrices we can write
$U^*_{(\bt)} \cU_{(\bt-1)} U_{(\bt)}^\dagger 
= U(0,\Ltr{\hat m}{\bt-1}{1} \vec i+ \Ltr{\hat m}{\bt-1}{ 3 } \vec k)$
(with $\Ltr{\hat m}{\bt-1}{}= \Ltr{n}{\bt-1}{}$) and then
get easily
\begin{align*}
\cos(4 \pi \Ltr{r}{\bt}{2}) 
&=
\frac{\Ltr{\hat m}{\bt-1}{3} }{ \Ltr{m}{\bt-1}{} }
,
\nonumber\\
\sin(4 \pi \Ltr{r}{\bt}{2}) 
&=
-\frac{\Ltr{\hat m}{\bt-1}{1} }{\Ltr{m}{\bt-1}{}}
.
\end{align*}
Since $-\oh\le \Ltr{r}{\bt}{2} < \oh$ the matrix $\Ltr{R}{\bt}{}$ is completely
determined up to a sign.
This explicit solution shows that $\Ltr{R}{\bt}{}$ is independent on $r$.

The second equation of eq.s (\ref{calU-to-hyperU}) can be used to fix
completely $U_{[\infty]}$ since it can be rewritten as
\begin{align}
 U_{[\infty]} 
&=
\Ltr{R}{\bt}{}^{-1} (U_{(\bt)} U_{(\bt-1)}^{\dagger}) (U_{(\bt)} U_{(\bt-1)}^{\dagger})^T \Ltr{R}{\bt}{}
.
\end{align}
All the previous eq.s can be solved since they are consistent with the various properties of the involved
matrices, i.e. $\cU^T=\cU$, $ U_{[0]}^T= U_{[0]}$ and
$ U_{[\infty]}^T= U_{[\infty]}$.
In particular this last property is valid only because of the choice
of ``gauge''  (\ref{gauge-n2}).

Let us now consider what happens to the tilded quantities associated
with $\partial \tilde\cZ$.
All the previous equations remain unchanged with the substitution of
the untilded quantities with the tilded ones.
In particular the given quantities $M_{(t)}$ and $\cU_{(t)}$ are
connected with the tilded ones by complex conjugation.
This means that $\tilde V$ is the same of $V^*$ up to a phase.
It also follows that $\tilde R = \pm R$ since $R$ is a real matrix.
Finally, $\tilde U_{[]}$ are the complex conjugate of the
corresponding $U_{[]}$.
The important consequence is then that ${\vec {\tilde n}}_{()}$ are
the parameters associated with the complex conjugate of $U_{[]}$, i.e
${\vec {\tilde n}}=(-n_1, +n_2, -n_3)$.
However this does not mean that there exists one $\tilde \cE_{\hypbas{0}  s}$ 
for each  $\cE_{\hypbas{0}  r}$ since the determination of the
possible solutions requires fixing all integers $k$s and $\tilde k$s.
In fact we can determine the parameters $k$s and $\tilde k$s of the
solutions  $\cE_{\hypbas{0}  r}$  and $\tilde \cE_{\hypbas{0}  s}$ by
requiring a finite classical action.

Finally, the fact that the solution for $R$ is unique up to a sign has as a
consequence for the way we write the conditions which can be used to
fix the coefficients $a$ and $b$.
In fact we write the global boundary conditions (\ref{cZ-global-bc})   as
\begin{align}
&Z(x_{t-1}, x_{t-1}) - Z(x_{t}, x_{t})
=
f_{(t-1)} -f_{(t)} 
\nonumber\\
&=
\int_{x_{t}; u\in H}^{x_{t-1}} du\, \partial_u Z_L(u)
+
\int_{x_{t}; \bu\in H^-}^{x_{t-1}} d\bu\, \bar\partial_\bu Z_R(\bu)
\nonumber\\
&=
\int_{x_{t}; u\in H}^{x_{t-1}} du\, \partial_z \cZ_{(\bt)}(z)|_{z=u}
+
\cU_{(\bt)}
\int_{x_{t}; \bu\in H^-}^{x_{t-1}} d\bu\, \bar\partial_z
\cbZ_{(\bt)}(z)|_{z=\bu}
.
\end{align}
Now $\cU_{(\bt)}$ and $\tilde V_{(\bt)}=V^*_{(\bt)}$ conspire to allow to write
\begin{align}
&Z(x_{t-1}, x_{t-1}) - Z(x_{t}, x_{t})
=
f_{(t-1)} -f_{(t)} 
\nonumber\\
&=
U^\dagger_{(\bt)} \Ltr{R}{\bt}{}\Big[
\sum_r a_r  I_{L\, r\, (t)}
+
\sum_s b_s I_{R\, s\, (t)}
\Big]
,
\label{global-constraint-final}
\end{align}
where we have defined the coefficients
\begin{align}
I^i_{L\, r\, (t)}
&=
\int_{\omegat_{t}; \omegat\in H}^{\omegat_{t-1}} d\omegat\,  
\cE^i_{\hypbas{0} r}(\omegat)
\nonumber\\
I^i_{R\, s\, (t)}
&=
\int_{\omegat_{t}; \bomegat\in H^-}^{\omegat_{t-1}} d\bomegat\, 
\tilde \cE^i_{\hypbas{0} s}(\bomegat)
.
\label{IL-IR}
\end{align}
In the previous expression (\ref{global-constraint-final}) 
we can take $a,b\in \R$ and  we are not obliged to consider the
previous expression with $a\rightarrow |a|$ and $b \rightarrow |b|$ as
it would follow from the direct application of eq. (\ref{Vr-ansatz})
where the phase would cancel the corresponding phase of $a$ (and
similarly for $b$).

Finally notice that the previous equation is simply asserting  that
there exists well adapted coordinates where the computations are more straightforward.
In facts when we have decided which branes $\Ltr{D}{\bt}{}$ to use for the doubling
trick and then we have mapped in the proper way the original
worldsheet coordinate $u$ into $\omegat$ as in eq. (\ref{omegabt})
then the space coordinates 
\begin{equation}
Z_{\{F_{(\bt-1)}\}}
=  \Ltr{R}{\bt}{}^T U_{(\bt)} Z 
=  \Ltr{R}{\bt}{}^T  Z_{(D_\bt)} 
\end{equation}
are good local coordinates at the interaction point $x_{\bt-1}$,
i.e.
coordinates for which
the monodromy at $\omega_{(\bt) \bt-1}=0$ and  are specially well
suited to perform the explicit computations.

\subsection{The classical solution with $SU(2)$ monodromy}
We are now ready to compute the classical solution in the simplest of
all non abelian cases, i.e. when the global monodromy group is
$SU(2)$.
We leave for future publications more complex cases.
As we have discussed before in section \ref{sect:a_first_naive_look}
there is only one solution of the hypergeometric which is needed for  $\dz\cZ_{(\bt)}(z)$
\begin{align}
\cE_{\hypbas{0} r=1}(\omegat)=
\PPom
{0}{ -\bn_{(\bt-1)} }{-n_{(\bt-1)}}
{1}{-\bn_{(\bt+1)} }{-n_{(\bt+1)}}
{\infty}{2-\bn_{(\bt)}}{2-n_{(\bt)}}
\label{cE-Psymb}
\end{align}
and only one solution of the hypergeometric for $\dz\cbZ_{(\bt)}(z)$
\begin{align}
\tilde \cE_{\hypbas{0} s=1}(\omegat)=
\PPom
{0}{ -\bn_{(\bt-1)} }{-n_{(\bt-1)}}
{1}{-\bn_{(\bt+1)} }{-n_{(\bt+1)}}
{\infty}{2-\bn_{(\bt)}}{2-n_{(\bt)}}
\label{ctE-Psymb}
.
\end{align}
For determining the values of the constants $a, b, c, d$ and $f$ we
have to make use of the discussion done in section
\ref{sect:from_U2_monod_to_Papp_eq}.
Would we not we could make many different associations between the
previous $P$ symbols and  the $P$ symbol in eq. (\ref{Papperitz-Riemann-U2})
because of the permutation property of the $P$ symbol
(\ref{P-perm-prop}).
Moreover we would miss the proper normalizations.

Now comparing the previous $P$ symbols with the $P$ symbol in
eq. (\ref{Papperitz-Riemann-U2}) and using eq.s (\ref{u2-to-abcdf}) we
get the following values
\begin{align}
a
&=
n_{(\bt-1) } + n_{(\bt)} + n_{(\bt+1) } -1
\nonumber\\
b
&=
n_{(\bt-1) } - n_{(\bt)} + n_{(\bt+1) } 
\nonumber\\
c
&=
2 n_{(\bt-1) } 
\nonumber\\
d
&=
n_{(\bt-1)}  -1
\nonumber\\
f
&=
n_{(\bt-1)} + n_{(\bt+1) } -2
,
\end{align}
since we can identify
\begin{equation}
n_{\hyppar{0} }= n_{(\bt-1)},~~ 
n_{\hyppar{1} }= n_{(\bt+1)},~~
n_{\hyppar{\infty} }= n_{(\bt)}
.  
\end{equation}

For the the $\tilde \cE$ solution we have to remember that 
$\tilde n_{(\bt-1) 3}= -n_{(\bt-1) 3}$ and $\tilde n_{(\bt) 1}=
-n_{(\bt) 1}$  which stems from $\tilde M= M^*$.
Therefore we get
\begin{align}
\tilde a
&=
-n_{(\bt-1) } + n_{(\bt)} + n_{(\bt+1) } 
&& = a +1 -c
\nonumber\\
\tilde b
&=
-n_{(\bt-1) } - n_{(\bt)} + n_{(\bt+1) } +1
&& = b +1 -c
\nonumber\\
\tilde c
&=
2- 2 n_{(\bt-1) } 
&& = 2 -c
\nonumber\\
\tilde d
&=
- n_{(\bt-1)}
&&
\nonumber\\
\tilde f
&=
-n_{(\bt-1)} + n_{(\bt+1) } -1
.
\end{align}
We are now ready to write the $\cE_{\hypbas{0}}$ and 
$\tilde \cE_{\hypbas{0}}$ as
\begin{align}
\cE_{\hypbas{0} r=1}(\omega)
=
&
(-\omega)^{n_{(\bt-1) } -1}
(1-\omega)^{n_{(\bt+1) } -1}
\times
\nonumber\\
&
\vect
{
\FF
{n_{(\bt-1) } + n_{(\bt)} + n_{(\bt+1) } -1}
{n_{(\bt-1) } - n_{(\bt)} + n_{(\bt+1) } }
{2 n_{(\bt-1) }}
{\omega}
}{
\cN
(-\omega)^{1-2 n_{(\bt-1) }}
\FF
{-n_{(\bt-1) } + n_{(\bt)} + n_{(\bt+1) } }
{-n_{(\bt-1) } - n_{(\bt)} + n_{(\bt+1) } +1 }
{2-2 n_{(\bt-1) }}
{\omega}
}
,
\end{align}
and
\begin{align}
\tilde \cE_{\hypbas{0} s=1}(\omega)
=
&
\cN^{-1} \, \mat{0}{1}{-1}{0} \, \cE_{\hypbas{0} r=1}(\omega)
\end{align}
with\footnote{
From the fact that the quantity under square root is positive along
with 
$|n_{(\bt-1) } - n_{(\bt)}| < n_{(\bt+1) } < n_{(\bt-1) } + n_{(\bt)}$
which follows from $U(2)$ multiplication law
we deduce that 
$\sin[\pi (n_{(\bt-1) } + n_{(\bt)} - n_{(\bt+1) })] 
     \sin[\pi (n_{(\bt-1) } + n_{(\bt)} + n_{(\bt+1) })] >0$
and hence the $sign(\dots)$ in eq. (\ref{d0-ratio}) is positive.
} 
\begin{equation}
\cN=
- sign(n_{(\bt) 1})
 \sqrt{
  \frac{ \sin[\pi (n_{(\bt-1) } + n_{(\bt)} - n_{(\bt+1) })] 
   }{
     \sin[\pi (n_{(\bt-1) } + n_{(\bt)} + n_{(\bt+1) })] 
   }
  \frac{ \sin[\pi (- (n_{(\bt-1) } - n_{(\bt)}) + n_{(\bt+1) })] 
   }{
     \sin[\pi ( (n_{(\bt-1) } - n_{(\bt)}) + n_{(\bt+1) })] 
   }
 }
.
\end{equation}
Notice that  $\cE_{\hypbas{0} r=1}(\omega)$ and
$\tilde \cE_{\hypbas{0} s=1}(\omega)$ differ in nuce by the $\sigma_2$
matrix because for any $SU(2)$ matrix we have $U^*= \sigma_2 U \sigma_2$.

Because ot this we reabsorb $\cN$ in the definition of
$\cE_{\hypbas{0} r=1}(\omega)$ and from now on we use
\begin{align}
\tilde \cE_{\hypbas{0} s=1}(\omega)
=
&
i \sigma_2 \, \cE_{\hypbas{0} r=1}(\omega)
.
\label{tildecE-cE}
\end{align}

Then we can write
\begin{align}
\dz \cZ_{(\bt) r=1} &= a_{r=1} \cE_{\hypbas{0} r=1}(\omega)
\nonumber\\
\dz \cbZ_{(\bt) s=1} &= b_{s=1} \tilde \cE_{\hypbas{0} s=1}(\omega)
.
\end{align}

We are left with the task of fixing the two real coefficients $a_{r=1}$ and
$b_{s=1}$. 
This is done using any $t$ in the equation (\ref{global-constraint-final}).
All of them are equivalent and must be consistent because the counting
of the dof.s
For example we can write
\begin{align}
\Ltr{R}{\bt}{}^T U_{(\bt)}
(
f_{(\bt+1)} -f_{(\bt-1)} 
)
=
a_{r=1} \int_{0; \omega\in H}^{1} d\omega\, \cE_{\hypbas{0} r=1}(\omega)
+
b_{s=1} \int_{0; \bomega\in H^-}^{1} d\bomega\, \cE_{\hypbas{0}
  r=1}(\bomega).
\end{align}
One can in principle doubt that this system is solvable since 
we have two real unknowns $a$ and $b$ 
while we have to match four real numbers $f_{(t)}- f_{(t-1)}\in \C^2$. 
In order to see whether it is  consistent and solvable we can
however consider the following case 
\begin{align}
\Ltr{R}{\bt}{}^T U_{(\bt)}
(
f_{(\bt-1)} -f_{(\bt)} 
)
=
a_{r=1} \int_{-\infty}^{0} d\omega\, \cE_{\hypbas{0} r=1}(\omega)
+
b_{s=1} \int_{-\infty }^{0} d\omega\, \tilde \cE_{\hypbas{0}  s=1}(\omega)
,
\end{align}
where we do not need to distinguish whether we are integrating in the
upper or lower half plane since we are working in the principal sheet
in a region where the solution has not any cut which are those
depicted in figure \ref{fig:Connect_two_monodromies_1}.
In this case the rhs is real. 
Therefore we need to verify that also the lhs is real.
This is however true because of eq. (\ref{f-constraints}).

The integrals in the previous equation can be expressed with the help
of generalized hypergeometric functions as
\begin{align}
 \int_{-\infty}^{0} d \omega\, 
&
(-\omega)^{\alpha-1} (1-\omega)^{\beta-1}
\FF{a}{b}{c}{\omega}
\nonumber\\
&=
\frac{
\Gamma(c-a)\,\Gamma(b)\,\Gamma(\alpha)
}{
\Gamma(c)\,\Gamma(b-\beta+1)
}
\,\Gamma(b-\beta-\alpha+1)
~{}_3F_{2}(c-a, b, \alpha~;~ c,b-\beta+1~;~1)
.
\end{align}
Explicitly we find
\begin{align}
&\vect{
a_{r=1} \,I^1_{L\, 1\, (\bt)} +b_{s=1} \,I^2_{L\, 1\, (\bt)}
}{
a_{r=1} \,I^2_{L\, 1\, (\bt)} +b_{s=1} \,I^1_{L\, 1\, (\bt)}
}
=
\Ltr{R}{\bt}{}^T U_{(\bt)}
(
f_{(\bt-1)} -f_{(\bt)} 
)
,
\label{eq-fixing-N3-sol}
\end{align}
where in accordance with eq. (\ref{IL-IR}) we have defined
\begin{align}
I^1_{L\, 1\,(\bt)}
=&
\frac{
  \Gamma(n_{(\bt-1) } - n_{(\bt+1) } -n_{(\bt)} + 1)
  \,\Gamma(n_{(\bt-1) } + n_{(\bt+1) } -n_{(\bt)} )
  \,\Gamma(n_{(\bt-1) })
}{
  \Gamma(2 n_{(\bt-1) })
  \,\Gamma(n_{(\bt-1) } -n_{(\bt)} +1)
}
\,\Gamma(-n_{(\bt)} +1)
\nonumber\\
~{}_3F_{2}&(n_{(\bt-1) } - n_{(\bt+1) } -n_{(\bt)} + 1,~
n_{(\bt-1) } + n_{(\bt+1) } -n_{(\bt)},~
n_{(\bt-1) }
   ~;~ 
2 n_{(\bt-1) },~
n_{(\bt-1) } -n_{(\bt)} +1
   ~;~1)
\nonumber\\
I^2_{L\, 1\,(\bt)}
=&
\cN
\frac{
  \Gamma(-n_{(\bt-1) } - n_{(\bt+1) } -n_{(\bt)} + 2)
  \,\Gamma(-n_{(\bt-1) } + n_{(\bt+1) } -n_{(\bt)} +1)
  \,\Gamma(-n_{(\bt-1) }+1)
}{
  \Gamma(-2 n_{(\bt-1) }+2)
  \,\Gamma(-n_{(\bt-1) } -n_{(\bt)} +2)
}
\,\Gamma(-n_{(\bt)} +1)
\nonumber\\
~{}_3F_{2}&(-n_{(\bt-1) } - n_{(\bt+1) } -n_{(\bt)} + 2,~ 
-n_{(\bt-1) } + n_{(\bt+1) } -n_{(\bt)}+1,~
-n_{(\bt-1) }+1
   ~;~ 
\nonumber\\
&\phantom{(-n_{(\bt-1) } - n_{(\bt+1) } -n_{(\bt)} + 2, 
-n_{(\bt-1) } + n_{(\bt+1) },)}
-2 n_{(\bt-1) }+2,~
-n_{(\bt-1) } -n_{(\bt)} +2
   ~;~1)
\nonumber\\
=&
\cN
\frac{
  \Gamma(\bar n_{(\bt-1) } - n_{(\bt+1) } -n_{(\bt)} + 1)
  \,\Gamma(\bar n_{(\bt-1) } + n_{(\bt+1) } -n_{(\bt)} )
  \,\Gamma(\bar n_{(\bt-1) })
}{
  \Gamma(2 \bar n_{(\bt-1) })
  \,\Gamma(\bar n_{(\bt-1) } -n_{(\bt)} +1)
}
\,\Gamma(-n_{(\bt)} +1)
\nonumber\\
&
~{}_3F_{2}(\bar n_{(\bt-1) } - n_{(\bt+1) } -n_{(\bt)} + 1,
\bar n_{(\bt-1) } + n_{(\bt+1) } -n_{(\bt)},
\bar n_{(\bt-1) }
   ~;~ 
2 \bar n_{(\bt-1) },
\bar n_{(\bt-1) } -n_{(\bt)} +1
   ~;~1)
\label{I1L-I2L}
\end{align}

\section{The classical action}
\label{sect:the_classical_action}
The next task is to compute the classical action of the classical
configuration we have found.
Many of the recent papers use to express the classical action using
the KLT formalism, i.e. they express the double integral on the
complex plane as a sum of products of two line integrals.
In the abelian case win $\NB=3$ the final answer is then simply the
area of a triangle.
In view of the fact that we expect the classical action must have
something to do with an area we should suspect that an easier approach
should be available.
In facts it turns out that using the eom in the case of holomorphic
solutions, which is unfortunately not our case, 
we show that the classical action
can be easily expressed using the embedding data.

We start from the classical action (\ref{Classical-action}) and we use
the eom along with the finitness of the action to immediately write
\begin{align}
4 \pi\alpha' S_{E}^{(classical)}
&=
\int_H d x\, d y \left( (\partial_x X^I)^2 + (\partial_y X^I)^2
\right)
\nonumber\\
&=
-\int_{-\infty}^{\infty} d x\,  X^I \partial_y X^I |_{y=i 0^+}
\end{align}
then we use the existence of different boundary conditions and the
boundary conditions (\ref{bc-real-good-coords}) along with
(\ref{X-loc-X-glo}) to write
\begin{align}
&=
-\sum_{t=1}^{\NB}
\int_{x_t}^{x_{t-1}} d x\,  X^I \partial_y X^I |_{y=i 0^+}
\nonumber\\
&=
-\sum_{t=1}^{\NB}
g^{N_{(t)}}_{(t)}
\int_{x_t}^{x_{t-1}} d x\,   \partial_y X^{N_{(t)}}_{(D_t)} |_{y=i 0^+}
\nonumber\\
&=
-i \sum_{t=1}^{\NB}
g^{N_{(t)}}_{(t)} R^{N_{(t)}}_{(t)~ J}
\int_{x_t}^{x_{t-1}} d x\,   ( X^{J '}_L(x +i 0^+) - X^{J '}_R(x -i 0^+) )
\nonumber\\
&=
2 \sum_{t=1}^{\NB}
g^{N_{(t)}}_{(t)}\, R^{N_{(t)}}_{(t)~ J}\,
\Im ( X^{J}_L(x_{t-1} +i 0^+)  - X^{J}_L(x_{t} +i 0^+)  )
.
\end{align}
This expression can be written using complex coordinates as
\COMMENTO{ 
Right expression but non in the desired form
\begin{align}
&=
-2 \sum_{t=1}^{\NB}
\Im \left[ g^{i}_{(t)} \right] 
\Re \Big[
&& 
U^{i}_{(t)~ j} 
\left( Z^j_L(x_{t-1} +i 0^+) -Z^j_L(x_{t} +i 0^+) \right)
\nonumber\\
& &&-
\left( U^{i}_{(t)~ j} 
  \left( Z^j_R(x_{t-1} -i 0^+) -Z^j_R(x_{t} -i 0^+) \right)  
\right)^*
\Big]
\label{Sclass-cpl-coords}
\end{align}
}
\begin{align}
&=
-2 \sum_{t=1}^{\NB}
\Im \left[ g^{i}_{(t)} \right] 
\left. \Re \left[ 
U^{i}_{(t)~ j}  Z^j_L(x +i 0^+)  
-
U^{i}_{(t)~ j}  Z^j_R(x -i 0^+) 
\right] \right|^{x=x_{t-1}}_{x=x_{t}}
\label{Sclass-cpl-coords}
\end{align}

\subsection{The holomorphic case}

\begin{figure}[hbt]
\begin{center}
\def\svgwidth{300px}
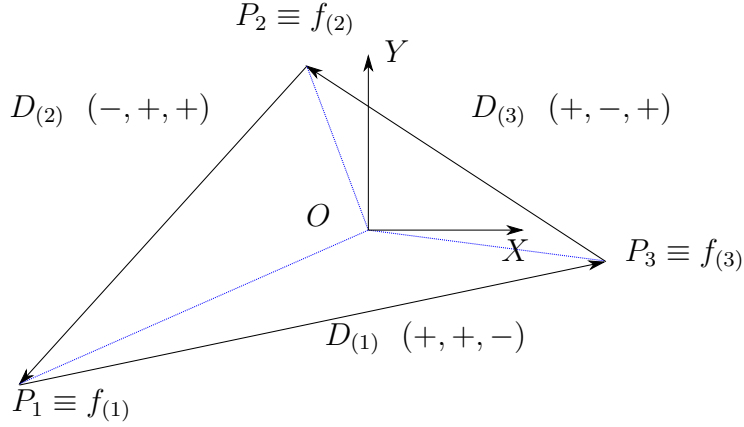
\end{center}
\vskip -0.5cm
\caption{
It is shown as the area of the triangle $ P_1 P_2 P_3$  is
decomposed into three subtriangles $P_1 P_3 O$, $P_3 P_2 O$ and $P_2
P_1 O$. 
Along each side the three signs are the signs of 
$\Re \left[ U^{z}_{(t)~ z} (f^z_{(t-1)} -f^z_{(t)}) \right]$, $a_{(t)}$ and
  $b_{(t)}$ where $Y=a_{(t)} X+ b _{(t)}$ is the equation of the line
  through the side.
}
\label{fig:R2_N3_area}
\end{figure}

\begin{figure}[hbt]
\begin{center}
\def\svgwidth{300px}
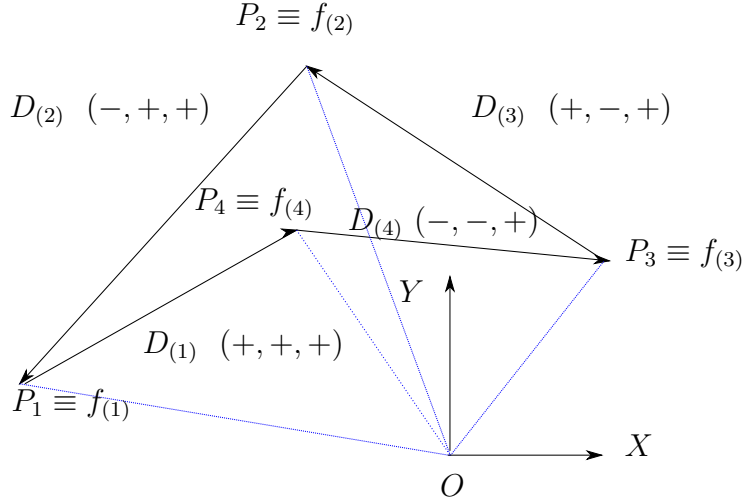
\end{center}
\vskip -0.5cm
\caption{
It is shown as the area of the polygon $ P_1 P_2 P_3 P_4$  is
decomposed into four subtriangles $P_1 P_4 O$, $P_4 P_3 O$ 
and $P_3 P_2 O$, $P_2 P_1 O$. 
The contributions to $S^{classical}_E$ 
of the first two triangles are positive while the
ones from the last two is negative thus yielding minus the area of the
interior of the polygon.
Along each side the three signs are the signs of 
$\Re \left[ U^{z}_{(t)~ z} (f^z_{(t-1)} -f^z_{(t)}) \right]$, $a_{(t)}$ and
  $b_{(t)}$ where $Y=a_{(t)} X+ b _{(t)}$ is the equation of the line
  through the side.
}
\label{fig:R2_N4_area}
\end{figure}

It is now immediate to finish the computation when $X^i$ is
holomorphic (or antiholomorphic) since in this case 
$X^j(x_t, \bar x_t)= X^j_L(x_t)= f^j_{(t)}$.
We therefore get
\begin{align}
4 \pi\alpha' S_{E}^{(classical)}
&=
-2 \sum_{t=1}^{\NB}
\Im\left[ g^{j}_{(t)} \right]\, 
\Re \left[ U^{i}_{(t)~ j} (f^j_{(t-1)} -f^j_{(t)}) \right]
.
\label{Sclas-holo}
\end{align}
In the case of $\R^2$ it is not difficult to see that the previous
expression is actually the area of the polygon bounding the string.
Few observations are needed to show this.
The quantity $ \sqrt{2}\, |\Im\left[ g^{z}_{(t)} \right]|  $ is the
distance of the side from the origin and 
$ \sqrt{2}\, |\Re \left[ U^{z}_{(t)~ z} (f^z_{(t-1)} -f^z_{(t)})
\right]|$ is the length of the side.
Given the equation of the line through the side $Y= a X + b$ the sign
of $\Im\left[ g^{z}_{(t)} \right]$ is the same of the sign of the
product $a b$.
The sign of $\Re \left[ U^{z}_{(t)~ z} (f^z_{(t-1)} -f^z_{(t)}) \right]$ is
  the same of the component of the vector $f^z_{(t-1)} -f^z_{(t)}$ on
  the $Y$ axis.
Then using the previous rules it is easy to see that each term of the
sum computes the (signed) area of a subtriangle of the original
polygon and that the sum is the area of the polygon
In figure (\ref{fig:R2_N3_area}) we give a very simple example 
of how this works and in figure (\ref{fig:R2_N4_area}) a less trivial one.
\COMMENTOO{
We deduce therefore that for any $t$ the expression 
$-2
\Im\left[ g^{j}_{(t)} \right]\, 
\Re \left[ U^{i}_{(t)~ j} (f^j_{(t-1)} -f^j_{(t)}) \right]
$
is the area with sign of the triangle determined having vertices 
$f_{(t-1)}$, $f_{(t)}$ and the origin.
This happens because this expression is invariant under rotation
$O(N)\subset U(N) \subset SO(2N)$ by which we can rotate the triangle
in a plane.
}

\subsection{The general case}

\COMMENTOO{
The action should be equivalent to Nambu-Goto action which
is the area of the minimal surface.
For $\R^2$  this is the area of the polygon BUT this is not what we
find in the non holomoprphic cases!!!
}

In the case where the solution is neither holomoprhic nor
antiholomorphic the computation is more complex and requires the
explicit knowledge of the solution.
To simplify the expression (\ref{Sclass-cpl-coords}) we start noticing that 
eq.s (\ref{Dt-bc}) imply that
\begin{equation}
\Im\left[ U^i_{(t)\,j}\, \partial_x Z^j_L(x)\right]
=\Im\left[ U^i_{(t)\,j}\, \partial_x Z^j_R(x)\right]
=0,
~~~~
x\in\R-\{x_t\}_{t=1,\dots\NB},
\label{Im-ZL-ZR-0}
\end{equation}
i.e.  both $\Im\left[ U_{(t)}\, Z_L(x)\right]$ and 
$\Im\left[ U_{(t)}\, Z_R(x)\right]$ are step functions with
discontinuities in the set of the interaction points $\{x_t\}_{t=1,\dots\NB}$.
In particular, from eq.s (\ref{Im-ZL-ZR-0}) it follows  that
$U^i_{(t)\,j}\, \partial_x Z^j_L(x)$ and
$U^i_{(t)\,j}\, \partial_x Z^j_R(x)$ are real vectors
hence eq. (\ref{Sclass-cpl-coords}) can be written without the real
projection as
\begin{align}
4 \pi\alpha' S_{E}^{(classical)}
&=
-2 \sum_{t=1}^{\NB}
\Im \left[ g^{i}_{(t)} \right] 
\left. U^{i}_{(t)~ j}  \left[  Z^j_L(x +i 0^+)  
-
Z^j_R(x -i 0^+) 
\right] \right|^{x=x_{t-1}}_{x=x_{t}}
.
\label{Sclass-cpl-coords-wo-real-proj}
\end{align}
The advantage of this expression is that we can directly compare with
the global boundary conditions which can be written as
\begin{equation}
\left. U^{i}_{(t)~ j}  \left[  Z^j_L(x +i 0^+)  
+
Z^j_R(x -i 0^+) 
\right] \right|_{x=x_{t}} 
= f^i_{(t)} 
,
\end{equation}
to see that once we have solved for the global contition we need not
to do more efforts to compute the classical action.

The classical action can be written also as
\begin{align}
4 \pi\alpha' S_{E}^{(classical)}
&=
-2 \sum_{t=1}^{\NB}
\Im \left[ g^{i}_{(t)} \right] 
 U^{i}_{(t)~ j}  \left[  
\left( f^i_{(t-1)} -f^i_{(t)} \right)
-
\left. 2 Z^j_R(x -i 0^+) \right|^{x=x_{t-1}}_{x=x_{t}} 
\right] 
.
%
\end{align}
in such a way to show the deviation from the holomorphic case and that
the real projection in eq. (\ref{Sclas-holo}) is not really necessary.
We can also make contact with the more explicit expression in
eq. (\ref{global-constraint-final}) by writing
\begin{align}
4 \pi\alpha' S_{E}^{(classical)}
&=
-2 \sum_{t=1}^{\NB}
\Im \left[ g^{i}_{(t)} \right] 
\left[  
 U^{i}_{(t)~ j}  \left( f^i_{(t-1)} -f^i_{(t)} \right)
-
 2 \left( U_{(t)} U_{(\bt)}^\dagger R_{(\bt)} \right)^{i}_{~ j}
\sum_s b_s I^j_{R\, s\, (t)}
\right] 
.
%
\end{align}

\noindent {\large {\bf Acknowledgments}}

This work is partially supported by the Compagnia di San Paolo
contract “MAST: Modern Applications of String Theory”
TO-Call3-2012-0088.

\appendix

\section{Fuchsian differential equations}
\label{app:fuchsian}
Fuchsian differential equations are linear differential equations
in the complex plane $P^1$ where all solutions near the singular
points of the coefficients are regular, i.e. their growth is bounded (in
any small sector) by an algebraic function.

We can therefore consider the linear differential equation of order
$n$ with meromorphic coefficients $p_i(z)$ ($i=1\dots n$) as
\begin{align}
\frac{ d^n y}{d z^n}
+
p_1(z) \frac{ d^{n-1} y}{d z^{n-1}}
+
p_2(z) \frac{ d^{n-2} y}{d z^{n-2}}
+
p_n(z) y
=0.
\end{align}
If we suppose that $z=z_i$ at finite is a singular point of the meromorphic
coefficients and we require to have $n$ solutions at this point with
behavior like $y\sim (z-z_i)^{\rho_i}$ we deduce immediately that the
previous equation can be written as
\begin{align}
\frac{ d^n y}{d z^n}
+
\frac{R_1(z-z_i)}{(z-z_i)} \frac{ d^{n-1} y}{d z^{n-1}}
+
\frac{R_2(z-z_i)}{(z-z_i)^2} \frac{ d^{n-2} y}{d z^{n-2}}
+ \dots+
\frac{R_n(z-z_i)}{(z-z_i)^n} y
=0
,
\end{align}
where $R_i$ are regular functions at $z=z_i$.
The possible values of $\rho_i$ are the indeces at $z=z_i$.

If we consider $N-1$ singular points at finite  we get therefore
\begin{align}
\frac{ d^n y}{d z^n}
+
\frac{\hat P_{1}(z)}{Q_{N-1}(z)} \frac{ d^{n-1} y}{d z^{n-1}}
+
\frac{\hat P_{2}(z)}{Q_{N-1}^2(z)} \frac{ d^{n-2} y}{d z^{n-2}}
+ \dots+
\frac{\hat P_{n}(z)}{Q_{N-1}^n(z)} y
=0
,
\end{align}
where $Q_{N-1}(z)=\prod_{j=1}^{N-1}(z-z_j)$ and $\hat P_i$ are
polynomials.

If we now want the infinity to be a regular singual point, i.e.
 we require to have $n$ solutions  with
behavior like $y\sim \left( \frac{1}{z} \right)^{\rho_\infty}$
we deduce that the most general Fuchsian equation with $N-1$ regular
singular points at finite and one at the infinite is given by 
\begin{align}
\frac{ d^n y}{d z^n}
+
\frac{P_{N-2}(z)}{Q_{N-1}(z)} \frac{ d^{n-1} y}{d z^{n-1}}
+
\frac{P_{2(N-2)}(z)}{Q_{N-1}^2(z)} \frac{ d^{n-2} y}{d z^{n-2}}
+ \dots+
\frac{P_{n(N-2)}(z)}{Q_{N-1}^n(z)} y
=0
\end{align}
where $Q_{N-1}(z)=\prod_{j=1}^{N-1}(z-z_j)$ when and $P_k(z)$ are abitrary
polynomials of order $k$.
We consider the case where the infinity is a singular point since it is
not immediate to write down the conditions for the regularity a infinity.

The previous equation for the special case $n=2$ can be written in a
more explicit form as
\begin{align}
\frac{ d^2 y}{d z^2}
+
\sum_{i=1}^{N-1} \frac{1- \rho_{i\, 1} \rho_{i\, 2}}{z - z_i}
\frac{ d y}{d z}
+
\sum_{i=1}^{N-1} \frac{\rho_{i\, 1}\rho_{i\, 2}+\gamma_i(z-z_i)}{(z - z_i)^2}
y=0
,
\end{align}
with $\sum_i \gamma_i=0$.

It is also easy to prove Fuchs' result according to which
the sum of all indeces must be equal to 
$(N-2) n (n-1)/2$, i.e.
\begin{equation}
\sum_{a=1}^n \left[
\sum_{i=1}^N \rho_{i\, a} + \rho_{\infty\, a}
\right]= (N-2) n (n-1)/2
.
\end{equation}
Given the behavior $y\sim (z-z_i)^{\rho_i}$ at the singular point
$z=z_i$ at finite
it is easy to show that the sum of all indeces is given by 
$\sum_{a=1}^n \rho_{i\, a}= n(n-1)/2-
P_{N-2}(z_i)/ [ \prod_{k\ne i} (z_i-z_k) ]$.
In fact we get near $z=z_i$
\begin{equation}
\rho_i(\rho_i-1) \dots (\rho_i-n+1) (z-z_i)^{\rho_i}+
\rho_i(\rho_i-1) \dots (\rho_i-n+2) 
\frac{ P_{N-2}(z_i)}{\prod_{j=1,    j\ne i}^{N-1}(z_i-z_j) }
(z-z_i)^{\rho_i}+ \dots =0
.
\end{equation}

Similarly assuming $y\sim \left( \frac{1}{z} \right)^{\rho_\infty}$ 
near $z=\infty$ we get
$ \sum_{a=1}^n \rho_{\infty\, a}= -n(n-1)/2 +A$ 
with $P_{N-2}(z)\sim A
z^{N-2}$ since
\begin{equation}
z^{-\rho_\infty+ n(N-2)} \left[
-\rho_\infty(-\rho_\infty-1) \dots (-\rho_\infty-n+1) 
-\rho_\infty(-\rho_\infty-1) \dots (-\rho_\infty-n+2) 
A
+ \dots \right] =0
.
\end{equation}

Using the residue theorem we get 
\begin{equation}
\oint_{|z|> max\{|z_i|\}} \frac{P_{N-2}(z)}{ Q_{N-1}(z)}\frac{ d z }{ 2 \pi i} 
= \sum_{i=1}^{N-2} \frac{  P_{N-2}(z_i) }{  \prod_{k\ne i} (z_i-z_k) }
=A
\end{equation}
from which the theorem follows.

Now a simple parameters counting gives $\sum_{i=1}^n [i (N-2)+1]$
parameters in the polynomials $P$ and $n N$ indeces of which only $n N-1$
are arbitrary by virtue of Fuchs' result.
Therefore we have  $N n(n-1)/2 - n^2 +1$ free accessory parameters. 
Only for $n=1$ and $n=2,~N=3$ there are no accesory parameters.

The general solution of the Fuchsian differential equation of order
$n$ with $N$ singularities can be represented by a generalized
$P$-symbol as
\begin{align}
y=
P\left\{ \begin{array}{c c c c c c} 
x_1 & x_2 & \dots & x_N & &\\   
\rho_{1\, 1} & \rho_{2\, 1} & \dots & \rho_{N\, 1} & \vec q &z\\   
\vdots & \vdots  & \dots & \vdots & & \\   
\rho_{1\, n} & \rho_{2\, n} & \dots & \rho_{N\, n} & & \\   
\end{array} \right\}
,
\end{align}
where $\vec q\in \C^{N n(n-1)/2 - n^2 +1}$.
This symbol represents the space of all $\infty^n$ solutions.

Since the equation is defined on $P^1$ the symbol is invariant under a $SL(2,\C$
 transformation $\hat z= (a z +b )/ (c z +d)$, i.e.
\begin{align}
y
=
P\left\{ \begin{array}{c c c c c c} 
x_1 & x_2 & \dots & x_N & &\\   
\rho_{1\, 1} & \rho_{2\, 1} & \dots & \rho_{N\, 1} & \vec q &z \\   
\vdots & \vdots  & \dots & \vdots & & \\   
\rho_{1\, n} & \rho_{2\, n} & \dots & \rho_{N\, n} & & \\   
\end{array} \right\}
=
P\left\{ \begin{array}{c c c c c c} 
\hat x_1 & \hat x_2 & \dots & \hat x_N & &\\   
\rho_{1\, 1} & \rho_{2\, 1} & \dots & \rho_{N\, 1} & \vec q & \hat z\\   
\vdots & \vdots  & \dots & \vdots & & \\   
\rho_{1\, n} & \rho_{2\, n} & \dots & \rho_{N\, n} & & \\   
\end{array} \right\}
.
\end{align}

\section{From discontinuities to monodromies}
\label{app:disc2mon}
In this appendix we would like to show explicitely in a local setup
that the monodromies depend on the base point.
This discussion is useful for the derivation of the monodromies
of the non abelian twists.

\begin{figure}[hbt]
\begin{minipage}[b]{0.45\linewidth}
\begin{center}
\def\svgwidth{0.8\linewidth}
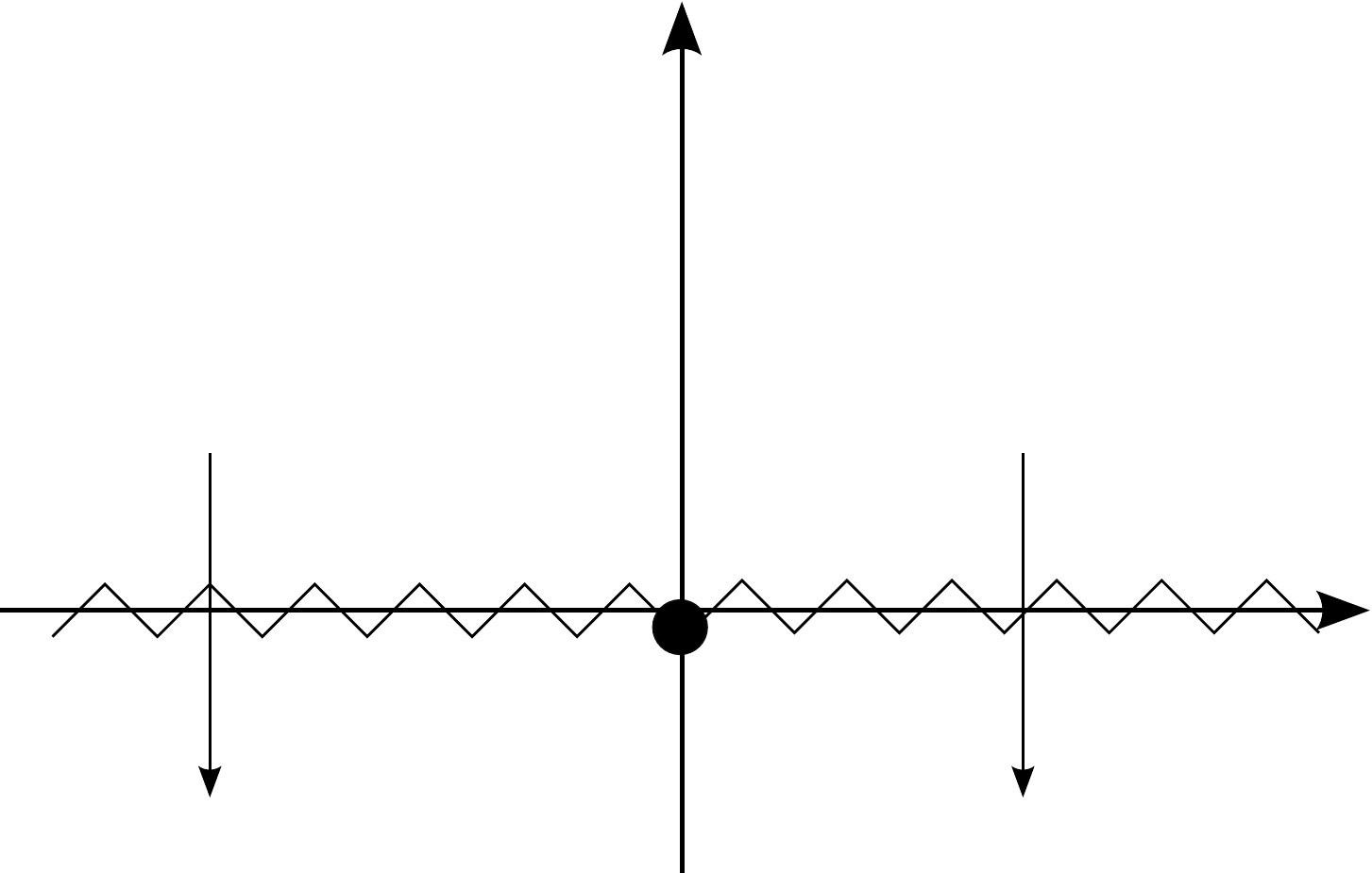
\end{center}
\vskip -0.5cm
\caption{The array of functions $f(z)$ with their discontinuities.
}
\label{fig:local_2_disc}
\end{minipage}
\begin{minipage}[b]{0.45\linewidth}
\begin{center}
\def\svgwidth{0.8\linewidth}
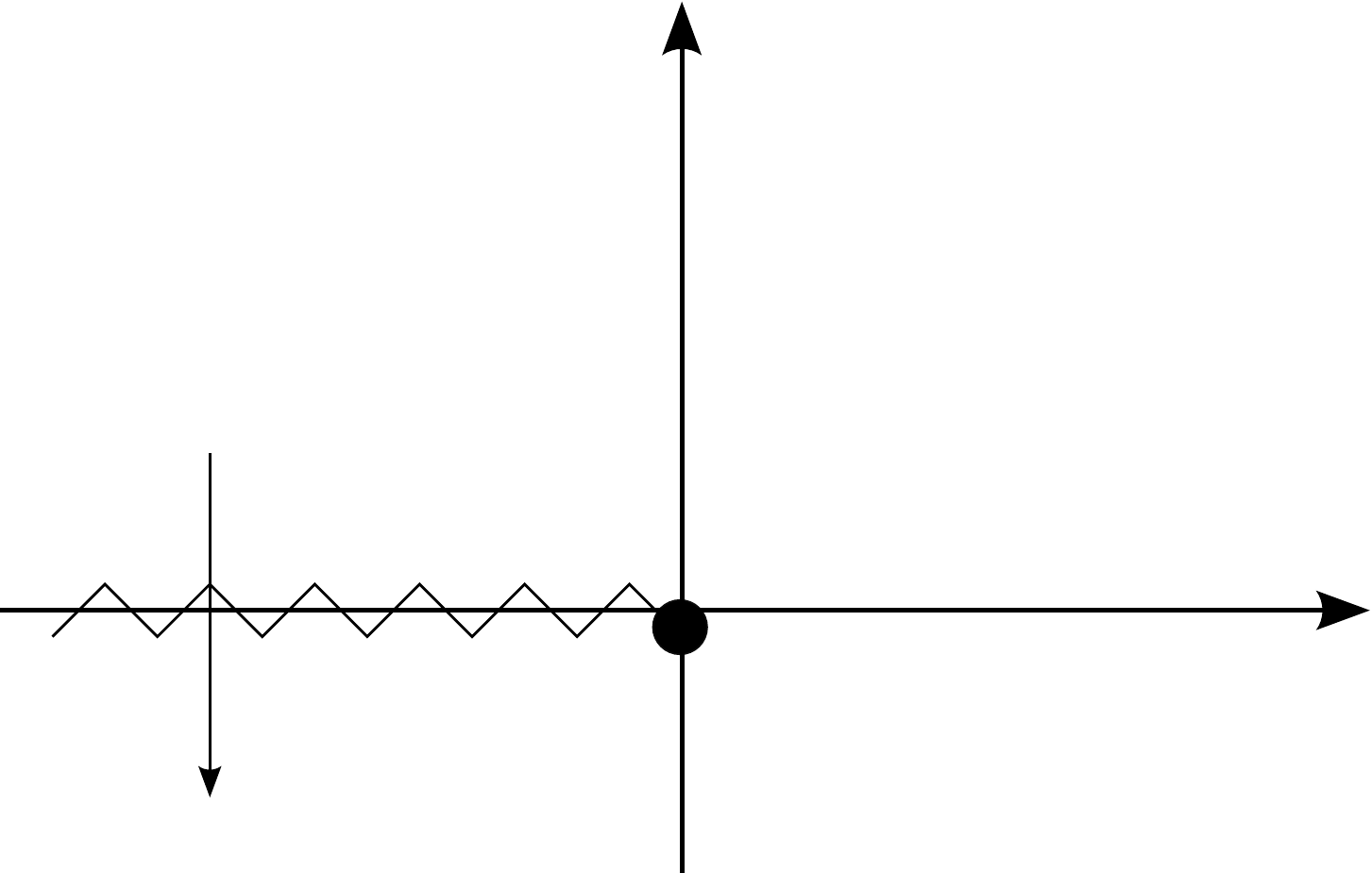
\end{center}
\vskip -0.5cm
\caption{The array of functions $F_+(z)$ with their discontinuity.
}
\label{fig:local_1_disc}
\end{minipage}
\end{figure}

Suppose we are given a vector of analytic functions $f(u)$ 
with discontinuities
\begin{align}
f(x- i 0^+) &= U_+  f(x + i 0^+) 
\nonumber\\
f(-x- i 0^+) &= U_-  f(-x + i 0^+)
,
\end{align}
with $x\in R$ and $x>0$ as shown in figure \ref{fig:local_2_disc}.

We want to compute the monodromies associated with 
$ f( e^{i 2 \pi} (x + i 0^+))$
and
$ f( e^{i 2 \pi} (x - i 0^+))$.
From figure \ref{fig:local_2_disc} it is clear that when we compute 
$ f( e^{i 2 \pi} (x + i0^+))$ 
we first cross the $U_-$ discontinuity and then the $U_+$
discontinuity in the opposite direction with respect to the
definition and therefore we have a contribution $U_+^{-1}$.
What it is not obvious is whether the final contribution is 
$U_-U_+^{-1}$ or $U_+^{-1} U_- $.
To solve this issue we define a new function $F_+(z)$ by the doubling
trick as
\begin{align}
F_+(z)
&=
\left\{\begin{array}{cc}
f(u) & z=u\in \mathring H \cup (0,+\infty) 
\\
U_+^{-1} f(\bu) & z=\bu\in \mathring H^- \cup (0,+\infty) 
\end{array}
\right.
.
\end{align}
This new function has only one discontinuity
\begin{equation}
F_+(-x- i 0^+) = U_+^{-1} U_-  F_+(-x + i 0^+)
,
\end{equation}
as shown in figure \ref{fig:local_1_disc}
It follows then that the monodromy is simply
\COMMENTOO{NON E' LA MATRICE ALLA -1??? SI' cambiato}
\begin{equation}
F_+\left(e^{i 2\pi} (x\pm i 0^+)\right) = U_-^{-1} U_+  F_+( x \pm i 0^+).
\end{equation}
In particular using the definition it follows that
\begin{align}
f\left(e^{i 2\pi} (x + i 0^+)\right) = U_-^{-1} U_+  f( x + i 0^+).
\nonumber\\
f\left(e^{i 2\pi} (x- i 0^+)\right) = U_+ U_-^{-1}   f( x - i 0^+).
\end{align}
This last results shows again that the monodromies depend on the base
point
and it is in accordance with the fact that to compute
$ f( e^{i 2 \pi} (x - i0^+))$ 
we first cross the $U_+^{-1}$ discontinuity in the opposite direction 
and then the $U_-$ discontinuity.

Finally, notice that the same result can be obtained usong a diferent
glueing, i.e.
\begin{align}
F_-(z)
&=
\left\{\begin{array}{cc}
f(u) & z=u\in \mathring H \cup (-\infty,0) 
\\
U_-^{-1} f(\bu) & z=\bu\in \mathring H^- \cup (-\infty,0) 
\end{array}
\right.
.
\end{align}

\section{Useful formula for $U(2)$}
\label{app:useful_U2_formula}
We parametrize a $U(2)$ matrix as
\COMMENTO{Check how the change of the $N$ definition 
range changes the mapping $U(2)$ params hypergeo ones.}
\begin{align}
U&= 
\exp(i\, 2\pi\, N ) \exp(i\, 2 \pi\, n^i \sigma_i)
\nonumber\\
&=e^{i\, 2\pi\, N}
\left( \cos(2\pi n)+ i \sin(2\pi n) \frac{\vec n}{n}
\cdot \vec \sigma\right),
\end{align}
with
\begin{align}
(N,\vec n)\in
\left\{-\frac{1}{4}\le N < \frac{1}{4},  0\le n \le \oh \right\}/ \sim
~~~~ 
(N,\vec n) \sim (N',\vec n') \mbox{ iff }
N=N', n=n'=\oh 
,
\label{canonical-U2-param}
\end{align}
where $\vec n= (n^1, n^2, n^3)$  and $n= |\vec n|$.

This  parametrization has the following two properties:
\begin{align}
-U(N,\vec n)
=U\left( N, -(\oh-|\vec n|) \frac{\vec n}{n}\right)
,
\end{align}
and
\begin{align}
[U(N,\vec n)]^* 
=U\left( -N, \vec {\tilde n} \right) 
= \sigma_2 U\left( -N, \vec n \right) \sigma_2
,
\end{align} 
for all $N$ with
${\vec {\tilde n}}= {(-n^1,+n^2,-n^3)}$.

It is also useful to record the product of two $U(2)$ elements. We
have in fact $U(M*N,\vec m*\vec n)= U(M,\vec m) U(N,\vec n)$ with
\begin{align}
U(N*M,  \vec m * \vec n)
=
\left\{
\begin{array}{c  l}
U(N+M+\oh, -(\oh - A) \frac{\vec A}{A} ) & -\oh\le N+M<-\frac{1}{4} \\
U(N+M, \vec A) & -\frac{1}{4}\le N+M<\frac{1}{4} \\
U(N+M-\oh, -(\oh - A) \frac{\vec A}{A} ) & -\frac{1}{4}\le N+M<-\frac{1}{2} 
\end{array}
\right.
,
\label{U2U2_product}
\end{align}
where the vector $\vec A$ with $0\le A \le \oh$ is defined
\begin{align}
\cos(2\pi A) 
&= 
\cos(2\pi m ) \cos(2\pi n ) 
- \sin(2\pi m ) \sin(2\pi n ) \frac{ \vec m \cdot \vec n }{ m n} 
\nonumber\\
\sin(2\pi A) 
\frac{\vec A}{ A}
&=
\cos(2\pi m ) \sin(2\pi n ) \frac{\vec n}{ n}
+\sin(2\pi m ) \cos(2\pi n ) \frac{\vec m}{ m}
- \sin(2\pi m ) \sin(2\pi n ) \frac{ \vec m \times \vec n }{ m n} 
.
\end{align}

Notice that it is also possible to choose a different range for $N$
$0\le N<\oh$ and all the formulas in this section remain unchanged in
form.
With this range the mapping between the parameters of a complex
conjugate $U(2)$ element and the corresponding ones of the usual
element is more complex,explicitly
\begin{equation}
[U(N,\vec n)]^* 
= U\left( \oh-N, -(\oh-|\vec n|) \frac{\vec {\tilde n}}{\tilde n} \right) 
= \sigma_2 
U\left( \oh-N, -(\oh-|\vec n|) \frac{\vec n}{n} \right) 
\sigma_2
\end{equation}
for $N \ne 0$
and
\begin{equation}
[U(N=0,\vec n)]^* 
=U(N=0,\vec {\tilde n}) 
= \sigma_2 U\left( N=0, \vec n \right) \sigma_2
\end{equation}
with $ \frac{\vec {\tilde n}}{\tilde n}= \frac{(-n^1,+n^2,-n^3)}{n}$.

In appendix \ref{app:mapping_two_ranges} we exam the exact mapping.


\section{Details on $U(2)$ monodromies}
\label{app:U2_monodromies}
In this appendix we would like to give the details of the derivation
of the relation between the indeces and $\frac{\dzero{2}}{\dzero{1}}$
and the parameters of the $U(2)$ monodromies.

Since the monodromy at $z=1$ is fixed once we have the monodromies at
$z=0$ and $z=\infty$ we will not deal with it.

Our strategy is first to find the constraints on the solution
parameters so that the monodromies matrices are in $U(2)$ and then
find the relation between them and the $U(2)$ parameters.

\subsection{Constraints from  the monodromy at $z=0$}
If we start from the basis of solutions $\cE_{\hypbas{0}}(z)$ in  eq. (\ref{Basis0}) 
the monodromy at $z=0$ 
\begin{equation}
\cE_{\hypbas{0}}(z e^{i 2 \pi}) = U_{\hypbas{0} \hyppar{0}}\, \cE_{\hypbas{0}}(z)
,~~~~ |z|<1
\end{equation}
becomes
\begin{equation}
U_{\hypbas{0} \hyppar{0}} 
= D_{\hypbas{0}} ^{-1} M_{\hypbas{0} \hyppar{0}} D_{\hypbas{0}}
,
\end{equation}
where 
\begin{equation}
M_{\hypbas{0} \hyppar{0}}
=
\mat{e^{i 2\pi d} }{}{}{e^{i 2\pi (d-c) }}
=e^{i 2\pi d} M_{(B) \hypbas{0} \hyppar{0}}
\end{equation}
is the monodromy of the Barnes basis at $z=0$ 
(\ref{Barnes-basis-0}) times $(-z)^d \, (1-z)^{f-d}$
around $z=0$  and
\begin{equation}
d_{\hypbas{0}}= \mat{\dzero{1}}{}{}{\dzero{2}}
\end{equation}
is the matrix used to rescale the two independent solutions of the 
Barnes basis at $z=0$ (\ref{Barnes-basis-0})  independently of each other.
Now requiring that 
$U_{\hypbas{0} \hyppar{0}} U_{\hypbas{0}  \hyppar{0}}^\dagger=\uno$ 
since $D_{\hypbas{0}}$
commutes with $M_{\hypbas{0} \hyppar{0}}$  implies immediately that
\begin{equation}
d,c\in\R.
\end{equation}

\subsection{Constraints from  the monodromy at $z=\infty$}
In an analogous way the monodromy at $z=\infty$ 
\begin{equation}
\cE_{\hypbas{0}}(z e^{-i 2 \pi}) = U_{\hypbas{0} \hyppar{\infty}}\, \cE_{\hypbas{0}}(z)
,~~~~ |z|>1
\end{equation}
reads
\begin{equation}
U_{\hypbas{0} \hyppar{\infty}} = 
e^{-i 2\pi f} D_{\hypbas{0}} ^{-1}  M_{\hypbas{0} \hyppar{\infty}}  D_{\hypbas{0}}
=e^{-i 2\pi f} D_{\hypbas{0}} ^{-1}  
 C M_{\hypbas{\infty}  \hyppar{\infty}} C^{-1} 
D_{\hypbas{0}},
\label{Uinf-M0inf-Minfinf}
\end{equation}
where  
$M_{\hypbas{0} \hyppar{\infty}}$ is the monodromy of the Barnes basis at
$z=0$ (\ref{Barnes-basis-0})
around $z=\infty$, 
$M_{\hypbas{\infty} \hyppar{\infty}}$ is the monodromy of the Barnes basis at
$z=\infty$ (\ref{Barnes-basis-inf})
around $z=\infty$ given in eq. (\ref{Barnes-monodromy-inf-inf}), $C$
is the matrix which connects the Barnes basis at $z=0$ and $z=\infty$
given in eq. (\ref{Barnes-basis-0-inf}), i.e.
\begin{equation}
C=
\frac{1}{\sin[\pi(b-a)] }
\mat
{  \sin[\pi(c-a)]   }
{ -\sin[\pi(c-b)] } 
{ -\sin[\pi a] } 
{ \sin[\pi b] }.
\end{equation} 
Because of the previous expression we check whether 
the matrix $U_{\hypbas{0} \hyppar{\infty}}$ is unitary by checking 
$U_{\hypbas{0} \hyppar{\infty}}^\dagger=U_{\hypbas{0} \hyppar{\infty}}^{-1}$.
This amounts to impose
\begin{equation}
M_{\hypbas{\infty} \hyppar{0}}^\dagger\, P = 
e^{-4\pi \Im(f)} P\, M_{\hypbas{\infty} \hyppar{0}}^{-1}
,
\end{equation}
with $P=C^\dagger  (D_{\hypbas{0}}  D^\dagger_{\hypbas{0}}) ^{-1} C$.
Since $P$ must be invertible we have $P_{11}, P_{22}>0$ therefore we
get
\begin{align}
\Im(a)=\Im(b)=\Im(f)
\end{align}
while the constraints involving $P_{12}, P_{21}$ imply
\begin{align}
P_{12}=0 
\rightarrow 
\left | \frac{\dzero{2} }{\dzero{1} } \right|^2
= 
- \frac{C_{21}^* C_{22}}{ C_{11}^* C_{12}}
=
-\frac{
\sin(\pi a^*) \sin(\pi b)
}{
\sin[\pi (a^*-c) ] \sin[\pi (b-c) ]
}.
\end{align}
This constraint can be implemented as
\begin{align}
-4 {
\sin(\pi a^*) \sin(\pi b)
}{
\sin[\pi (a^*-c) ] \sin[\pi (b-c) ]
}\in \R^+.
\end{align}
The imaginary part of the previous product must vanish and can be written as 
\begin{align}
 \sin(i 4\pi \Im(a))\,  
\sin(\pi c)\, 
\left( sin( \pi( 2 \Re(a)-c)) - sin( \pi( 2 \Re(b)-c)) \right)
\end{align}
therefore we get three cases
either $\Im(a)=\Im(b)=\Im(f)$ 
or $\Re(a)=\Re(b)~~~mod~1$ 
or $c=\Re(a)+\Re(b)~~~mod~1$.
The latter case is not  admissible since 
$\left | \frac{\dzero{2} }{\dzero{1} } \right|^2$ would be negative.
While the second gives an abelian monodromy.

\subsection{The monodromy at $z=\infty$}
We start by taking the trace of eq. (\ref{Uinf-M0inf-Minfinf}) from
which we immediately get
\begin{equation}
 e^{i\, 2\pi\, N_\hyppar{\infty}} \cos(2\pi n_\hyppar{\infty} )
= e^{i \pi (a+b-f)} \cos(\pi(a-b))
\end{equation}
which implies eq.s (\ref{Ni_ni}).

Now taking the imaginary part of $U_{\hypbas{0} \hyppar{\infty}\, 1 1}$  and using eq.s
(\ref{Ni_ni})  we find immediately the first of eq.s (\ref{ni3_ni12}),
i.e.
\begin{align}
\frac{n^3_{\hyppar{\infty}}}{n_{\hyppar{\infty}}}
&=
 (-)^{s_{n \hyppar{\infty}} +1 }
\frac{
\cos[ \pi(a+b-c) ] - \cos( \pi c ) \cos[\pi(a-b) ]
}{
\sin( \pi c ) \sin[\pi(a-b) ]
}
.
\end{align}

From the product 
$U_{\hypbas{0} \hyppar{\infty}\, 1 2} 
U_{\hypbas{0}  \hyppar{\infty}\,  2 1} =
|U_{\hypbas{0} \hyppar{\infty}\, 1 2} |^2
$  
it follows the constraint in eq. (\ref{-sin-prod-gr-zero})
\begin{equation}
- \sin(\pi a) \sin(\pi b) \sin[\pi (a-c)] \sin[\pi (b-c)] >0
.
\end{equation}

From the ratio 
$U_{\hypbas{0} \hyppar{\infty}\, 1 2} / 
U_{\hypbas{0}  \hyppar{\infty}\,  2 1} 
$  
and from the previous equation
it follows eq. (\ref{modulus_d02/d01})
\begin{equation}
\left| \frac{\dzero{2}}{\dzero{1}} \right|^2
=
- \frac{ \sin(\pi a) \sin(\pi b)}{ \sin[\pi (a-c)] \sin[\pi (b-c)]}
.
\end{equation}

Finally from $U_{\hypbas{0} \hyppar{\infty}\, 1 2}$ we get the second
equation in eq.s (\ref{ni3_ni12})
\begin{align}
\frac{n^1_{\hyppar{\infty}} + i\, n^2_{\hyppar{\infty}}}
{ 
n_{\hyppar{\infty}} 
}
=&
e^{-i 2 \pi \delta_{\hyppar{0}} }
(-)^{s_{n_\hyppar{\infty}} +1  }\, sign( \sin(\pi a) \sin(\pi b) )
\nonumber\\
&\frac{ 
\sqrt{-4 \sin(\pi a) \sin(\pi b) \sin[\pi (a-c) ] \sin[\pi (b-c)]
  } 
}{
 \sin(\pi c) \sin[\pi (a-b) ]
}
.
\end{align}

\section{Details on the mapping of the Papperitz-Riemann 
parameters for the two $U(2)$ parametrizations.}
\label{app:mapping_two_ranges}
An element of $U(2)$ can be represented as
\begin{align}
U&= 
\exp(i\, 2\pi\, N ) \exp(i\, 2 \pi\, n^i \sigma_i)
\nonumber\\
&=e^{i\, 2\pi\, N}
\left( \cos(2\pi n)+ i \sin(2\pi n) \frac{\vec n}{n}
\cdot \vec \sigma\right)
.
\end{align}
There are at least two obvious parametrizations, the one chosen in themain text
\begin{align}
\left\{-\frac{1}{4}\le N < \frac{1}{4},~  0\le n \le \oh 
~|~ n=n'=\oh \Rightarrow \vec n \equiv \vec n' \right\}
,
\end{align}
and a different one given by
\begin{align}
\left\{0\le \hat N < \frac{1}{2},~  0\le \hat n \le \oh 
~|~ \hat n=\hat n'=\oh \Rightarrow \vec{\hat n} \equiv \vec{\hat n'} \right\}
.
\end{align}
These two parametrization are connected by
\begin{align}
(\hat N, \vec{\hat n})
=\left\{
\begin{array}{c c}
( N,~ \vec{ n}) & 0\le N <\frac{1}{4} \\
( N+\oh,~ (n-\oh) \frac{\vec{ n}}{n} ) & -\frac{1}{4}\le N<0 \\
\end{array}
\right.
.
\end{align}

As a check of eq.s (\ref{u2-to-abcdf}) we can verify that the parameters of 
the Papperitz-Riemann symbols are the same in the two paremetrizations 
up to the Papperitz-Riemann symbols symmetries.

For example for $\frac{1}{4}<N_{\hyppar{0}}<\oh$ and 
$0\le N_{\hyppar{\infty}}<\frac{1}{4}$ we get
\begin{align}
\hat a=a,~\hat b= b,~\hat c=c,~\hat d=d,~\hat f=f
\end{align}
when we map
\begin{align}
\hat A&=\oh-A
\nonumber\\
(-)^{\hat s_A}&=-(-)^{s_A}
\nonumber\\
\hat k_a &= k_a+\oh(1+(-)^{s_A})
\nonumber\\
\hat k_b &= k_b+\oh(1+(-)^{s_A})
\nonumber\\
\hat k_c &= k_c+1
\nonumber\\
\hat k_d &= k_d
\nonumber\\
\hat k_f &= k_f +\oh(1-(-)^{s_A})
.
\end{align}
Notice that $\hat A=\oh-A$ is enforced by eq. (\ref{A-definition}).

Another example is given by $0\le N_{\hyppar{0}}<\frac{1}{4}$ 
and $\frac{1}{4}<N_{\hyppar{\infty}}<\oh$  for which we have
\begin{align}
\hat a= b,~\hat b= a,~\hat c=c,~\hat d=d,~\hat f=f
\end{align}
when we map
\begin{align}
\hat A&=\oh-A
\nonumber\\
(-)^{\hat s_A}&=-(-)^{s_A}
\nonumber\\
\hat k_a &= k_b -\oh(1-(-)^{s_A})
\nonumber\\
\hat k_b &= k_a+\oh(1-(-)^{s_A})
\nonumber\\
\hat k_c &= k_c
\nonumber\\
\hat k_d &= k_d
\nonumber\\
\hat k_f &= k_f +\oh(1-(-)^{s_A})
.
\end{align}


\end{document}